%
%
%
\font\ninerm=cmr9
\font\eightrm=cmr8
\font\sixrm=cmr6
\font\ninei=cmmi9
\font\eighti=cmmi8
\font\sixi=cmmi6
\skewchar\ninei='177 \skewchar\eighti='177 \skewchar\sixi='177
\font\ninesy=cmsy9
\font\eightsy=cmsy8
\font\sixsy=cmsy6
\skewchar\ninesy='60 \skewchar\eightsy='60 \skewchar\sixsy='60

\font\ninebf=cmbx9
\font\eightbf=cmbx8
\font\sixbf=cmbx6
\font\ninett=cmtt9
\font\eighttt=cmtt8
\hyphenchar\tentt=-1 
\hyphenchar\ninett=-1
\hyphenchar\eighttt=-1
\font\ninesl=cmsl9
\font\eightsl=cmsl8
\font\nineit=cmti9
\font\eightit=cmti8
\newskip\ttglue
\def\tenpoint{\def\rm{\fam0\tenrm}%
  \textfont0=\tenrm \scriptfont0=\sevenrm \scriptscriptfont0=\fiverm
  \textfont1=\teni \scriptfont1=\seveni \scriptscriptfont1=\fivei
  \textfont2=\tensy \scriptfont2=\sevensy \scriptscriptfont2=\fivesy
  \textfont3=\tenex \scriptfont3=\tenex \scriptscriptfont3=\tenex
  \def\it{\fam\itfam\tenit}%
  \textfont\itfam=\tenit
  \def\sl{\fam\slfam\tensl}%
  \textfont\slfam=\tensl
  \def\bf{\fam\bffam\tenbf}%
  \textfont\bffam=\tenbf \scriptfont\bffam=\sevenbf
   \scriptscriptfont\bffam=\fivebf
  \def\tt{\fam\ttfam\tentt}%
  \textfont\ttfam=\tentt
  \tt \ttglue=.5em plus.25em minus.15em
  \normalbaselineskip=12pt
  \let\sc=\eightrm
  \let\big=\tenbig
  \setbox\strutbox=\hbox{\vrule height8.5pt depth3.5pt width0pt}%
  \normalbaselines\rm}
\def\ninepoint{\def\rm{\fam0\ninerm}%
  \textfont0=\ninerm \scriptfont0=\sixrm \scriptscriptfont0=\fiverm
  \textfont1=\ninei \scriptfont1=\sixi \scriptscriptfont1=\fivei
  \textfont2=\ninesy \scriptfont2=\sixsy \scriptscriptfont2=\fivesy
  \textfont3=\tenex \scriptfont3=\tenex \scriptscriptfont3=\tenex
  \def\it{\fam\itfam\nineit}%
  \textfont\itfam=\nineit
  \def\sl{\fam\slfam\ninesl}%
  \textfont\slfam=\ninesl
  \def\bf{\fam\bffam\ninebf}%
  \textfont\bffam=\ninebf \scriptfont\bffam=\sixbf
   \scriptscriptfont\bffam=\fivebf
  \def\tt{\fam\ttfam\ninett}%
  \textfont\ttfam=\ninett
  \tt \ttglue=.5em plus.25em minus.15em
  \normalbaselineskip=10pt 
  \let\sc=\sevenrm
  \let\big=\ninebig
  \setbox\strutbox=\hbox{\vrule height8pt depth3pt width0pt}%
  \normalbaselines\rm}
\def\eightpoint{\def\rm{\fam0\eightrm}%
  \textfont0=\eightrm \scriptfont0=\sixrm \scriptscriptfont0=\fiverm
  \textfont1=\eighti \scriptfont1=\sixi \scriptscriptfont1=\fivei
  \textfont2=\eightsy \scriptfont2=\sixsy \scriptscriptfont2=\fivesy
  \textfont3=\tenex \scriptfont3=\tenex \scriptscriptfont3=\tenex
  \def\it{\fam\itfam\eightit}%
  \textfont\itfam=\eightit
  \def\sl{\fam\slfam\eightsl}%
  \textfont\slfam=\eightsl
  \def\bf{\fam\bffam\eightbf}%
  \textfont\bffam=\eightbf \scriptfont\bffam=\sixbf
   \scriptscriptfont\bffam=\fivebf
  \def\tt{\fam\ttfam\eighttt}%
  \textfont\ttfam=\eighttt
  \tt \ttglue=.5em plus.25em minus.15em
  \normalbaselineskip=9pt
  \let\sc=\sixrm
  \let\big=\eightbig
  \setbox\strutbox=\hbox{\vrule height7pt depth2pt width0pt}%
  \normalbaselines\rm}
%
\def\headtype{\ninepoint}                 
\def\abstracttype{\ninepoint}             
\def\captiontype{\ninepoint}              
\def\footnotetype{\ninepoint}             
\def\refit{\it}                           
\font\chaptitle=cmr10 at 11pt             
\rm                                       

%
%
\parindent=0.25in                         
\parskip=0pt                              
\baselineskip=12pt                        
\hsize=4.25truein                         
\vsize=7.445truein                        
\hoffset=1in                              
\voffset=-0.5in                           

\newskip\sectionskipamount                
\newskip\aftermainskipamount              
\newskip\subsecskipamount                 
\newskip\firstpageskipamount              
\newskip\capskipamount                    
\newskip\ackskipamount                    
\sectionskipamount=0.2in plus 0.09in
\aftermainskipamount=6pt plus 6pt         
\subsecskipamount=0.1in plus 0.04in
\firstpageskipamount=3pc
\capskipamount=0.1in
\ackskipamount=0.15in
\def\sectionskip{\vskip\sectionskipamount}
\def\aftermainskip{\vskip\aftermainskipamount}
\def\subsecskip{\vskip\subsecskipamount} 
\def\firstpageskip{\vskip\firstpageskipamount}
\def\capskip{\hskip\capskipamount}

%
%
\nopagenumbers                            
\newcount\firstpageno                     
\firstpageno=\pageno                      
\newcount\chapno                          

\def\rightheadline{\headtype\phantom{\folio}\hfil\runningtitletext\hfil\folio}
\def\leftheadline{\headtype\folio\hfil\runningauthortext\hfil\phantom{\folio}}
\headline={\ifnum\pageno=\firstpageno\hfil
           \else
              \ifdim\ht\topins=\vsize           
                 \ifdim\dp\topins=1sp \hfil     
                 \else
                     \ifodd\pageno\rightheadline\else\leftheadline\fi
                 \fi
              \else
                 \ifodd\pageno\rightheadline\else\leftheadline\fi
              \fi
           \fi}

\def\bottomnumber{\hss\tenrm[\folio]\hss}
\footline={\ifnum\pageno=\firstpageno\bottomnumber\else\hfil\fi}

%
%
%
%
\outer\def\mainsection#1
    {\vskip 0pt plus\smallskipamount\sectionskip
     \message{#1}\vbox{\noindent{\bf#1}}\nobreak\aftermainskip\noindent}
 
\outer\def\subsection#1
    {\vskip 0pt plus\smallskipamount\subsecskip
     \message{#1}\vbox{\noindent{\bf#1}}\nobreak\smallskip\nobreak\noindent}
 
\def\backup{\nobreak\vskip-\baselineskip\nobreak\vskip-\subsecskipamount\nobreak
}

\def\title#1{{\chaptitle\leftline{#1}}}
\def\name#1{\leftline{#1}}
\def\affiliation#1{\leftline{\it #1}}
\def\abstract#1{{\abstracttype \noindent #1 \smallskip\vskip .1in}}
\def\ref{\noindent \parshape2 0truein 4.25truein 0.25truein 4truein}
\def\caption{\noindent \captiontype
             \parshape=2 0truein 4.25truein .125truein 4.125truein}

\def\footnote#1{\edef\fspafac{\spacefactor\the\spacefactor}#1\fspafac
      \insert\footins\bgroup\footnotetype
      \interlinepenalty100 \let\par=\endgraf
        \leftskip=0pt \rightskip=0pt
        \splittopskip=10pt plus 1pt minus 1pt \floatingpenalty=20000
        \textindent{#1}\bgroup\strut\aftergroup\strut\egroup\let\next}
\skip\footins=12pt plus 2pt minus 4pt 
\dimen\footins=30pc 

%
%

\def\@{\spacefactor 1000}

\def\,{\pcomma} 
\def\pcomma{\relax\ifmmode\mskip\thinmuskip\else\thinspace\fi}

\def\oversim#1#2{\lower0.5ex\vbox{\baselineskip=0pt\lineskip=0.2ex
     \ialign{$\mathsurround=0pt #1\hfil##\hfil$\crcr#2\crcr\sim\crcr}}}
\def\simgt{\mathrel{\mathpalette\oversim>}}
\def\simlt{\mathrel{\mathpalette\oversim<}}

%
%
%
%
%
\catcode`\@=11\relax
\newwrite\@unused
\def\typeout#1{{\let\protect\string\immediate\write\@unused{#1}}}
\typeout{psfig: version 1.1}

%
%
\def\@nnil{\@nil}
\def\@empty{}
\def\@psdonoop#1\@@#2#3{}
\def\@psdo#1:=#2\do#3{\edef\@psdotmp{#2}\ifx\@psdotmp\@empty \else
    \expandafter\@psdoloop#2,\@nil,\@nil\@@#1{#3}\fi}
\def\@psdoloop#1,#2,#3\@@#4#5{\def#4{#1}\ifx #4\@nnil \else
       #5\def#4{#2}\ifx #4\@nnil \else#5\@ipsdoloop #3\@@#4{#5}\fi\fi}
\def\@ipsdoloop#1,#2\@@#3#4{\def#3{#1}\ifx #3\@nnil 
       \let\@nextwhile=\@psdonoop \else
      #4\relax\let\@nextwhile=\@ipsdoloop\fi\@nextwhile#2\@@#3{#4}}
\def\@tpsdo#1:=#2\do#3{\xdef\@psdotmp{#2}\ifx\@psdotmp\@empty \else
    \@tpsdoloop#2\@nil\@nil\@@#1{#3}\fi}
\def\@tpsdoloop#1#2\@@#3#4{\def#3{#1}\ifx #3\@nnil 
       \let\@nextwhile=\@psdonoop \else
      #4\relax\let\@nextwhile=\@tpsdoloop\fi\@nextwhile#2\@@#3{#4}}
\def\psdraft{
	\def\@psdraft{0}
}
\def\psfull{
	\def\@psdraft{100}
}
\psfull
\newif\if@prologfile
\newif\if@postlogfile
\newif\if@bbllx
\newif\if@bblly
\newif\if@bburx
\newif\if@bbury
\newif\if@height
\newif\if@width
\newif\if@rheight
\newif\if@rwidth
\newif\if@clip
\def\@p@@sclip#1{\@cliptrue}
\def\@p@@sfile#1{
		   \def\@p@sfile{#1}
}
\def\@p@@sfigure#1{\def\@p@sfile{#1}}
\def\@p@@sbbllx#1{
		\@bbllxtrue
		\dimen100=#1
		\edef\@p@sbbllx{\number\dimen100}
}
\def\@p@@sbblly#1{
		\@bbllytrue
		\dimen100=#1
		\edef\@p@sbblly{\number\dimen100}
}
\def\@p@@sbburx#1{
		\@bburxtrue
		\dimen100=#1
		\edef\@p@sbburx{\number\dimen100}
}
\def\@p@@sbbury#1{
		\@bburytrue
		\dimen100=#1
		\edef\@p@sbbury{\number\dimen100}
}
\def\@p@@sheight#1{
		\@heighttrue
		\dimen100=#1
   		\edef\@p@sheight{\number\dimen100}
}
\def\@p@@swidth#1{
		\@widthtrue
		\dimen100=#1
		\edef\@p@swidth{\number\dimen100}
}
\def\@p@@srheight#1{
		\@rheighttrue
		\dimen100=#1
		\edef\@p@srheight{\number\dimen100}
}
\def\@p@@srwidth#1{
		\@rwidthtrue
		\dimen100=#1
		\edef\@p@srwidth{\number\dimen100}
}
\def\@p@@sprolog#1{\@prologfiletrue\def\@prologfileval{#1}}
\def\@p@@spostlog#1{\@postlogfiletrue\def\@postlogfileval{#1}}
\def\@cs@name#1{\csname #1\endcsname}
\def\@setparms#1=#2,{\@cs@name{@p@@s#1}{#2}}
%
%
\def\ps@init@parms{
		\@bbllxfalse \@bbllyfalse
		\@bburxfalse \@bburyfalse
		\@heightfalse \@widthfalse
		\@rheightfalse \@rwidthfalse
		\def\@p@sbbllx{}\def\@p@sbblly{}
		\def\@p@sbburx{}\def\@p@sbbury{}
		\def\@p@sheight{}\def\@p@swidth{}
		\def\@p@srheight{}\def\@p@srwidth{}
		\def\@p@sfile{}
		\def\@p@scost{10}
		\def\@sc{}
		\@prologfilefalse
		\@postlogfilefalse
		\@clipfalse
}
%
%
\def\parse@ps@parms#1{
	 	\@psdo\@psfiga:=#1\do
		   {\expandafter\@setparms\@psfiga,}}
%
%
\newif\ifno@bb
\newif\ifnot@eof
\newread\ps@stream
\def\bb@missing{
	\typeout{psfig: searching \@p@sfile \space  for bounding box}
	\openin\ps@stream=\@p@sfile
	\no@bbtrue
	\not@eoftrue
	\catcode`\%=12
	\loop
		\read\ps@stream to \line@in
		\global\toks200=\expandafter{\line@in}
		\ifeof\ps@stream \not@eoffalse \fi
		\@bbtest{\toks200}
		\if@bbmatch\not@eoffalse\expandafter\bb@cull\the\toks200\fi
	\ifnot@eof \repeat
	\catcode`\%=14
}	
\catcode`\%=12
\newif\if@bbmatch
\def\@bbtest#1{\expandafter\@a@\the#1
\long\def\@a@#1
\long\def\bb@cull#1 #2 #3 #4 #5 {
	\dimen100=#2 bp\edef\@p@sbbllx{\number\dimen100}
	\dimen100=#3 bp\edef\@p@sbblly{\number\dimen100}
	\dimen100=#4 bp\edef\@p@sbburx{\number\dimen100}
	\dimen100=#5 bp\edef\@p@sbbury{\number\dimen100}
	\no@bbfalse
}
\catcode`\%=14
\def\compute@bb{
		\no@bbfalse
		\if@bbllx \else \no@bbtrue \fi
		\if@bblly \else \no@bbtrue \fi
		\if@bburx \else \no@bbtrue \fi
		\if@bbury \else \no@bbtrue \fi
		\ifno@bb \bb@missing \fi
		\ifno@bb \typeout{FATAL ERROR: no bb supplied or found}
			\no-bb-error
		\fi
		\count203=\@p@sbburx
		\count204=\@p@sbbury
		\advance\count203 by -\@p@sbbllx
		\advance\count204 by -\@p@sbblly
		\edef\@bbw{\number\count203}
		\edef\@bbh{\number\count204}
}
%
%
\def\in@hundreds#1#2#3{\count240=#2 \count241=#3
		     \count100=\count240	
		     \divide\count100 by \count241
		     \count101=\count100
		     \multiply\count101 by \count241
		     \advance\count240 by -\count101
		     \multiply\count240 by 10
		     \count101=\count240	
		     \divide\count101 by \count241
		     \count102=\count101
		     \multiply\count102 by \count241
		     \advance\count240 by -\count102
		     \multiply\count240 by 10
		     \count102=\count240	
		     \divide\count102 by \count241
		     \count200=#1\count205=0
		     \count201=\count200
			\multiply\count201 by \count100
		 	\advance\count205 by \count201
		     \count201=\count200
			\divide\count201 by 10
			\multiply\count201 by \count101
			\advance\count205 by \count201
		     \count201=\count200
			\divide\count201 by 100
			\multiply\count201 by \count102
			\advance\count205 by \count201
		     \edef\@result{\number\count205}
}
\def\compute@wfromh{
		\in@hundreds{\@p@sheight}{\@bbw}{\@bbh}
		\edef\@p@swidth{\@result}
}
\def\compute@hfromw{
		\in@hundreds{\@p@swidth}{\@bbh}{\@bbw}
		\edef\@p@sheight{\@result}
}
\def\compute@handw{
		\if@height 
			\if@width
			\else
				\compute@wfromh
			\fi
		\else 
			\if@width
				\compute@hfromw
			\else
				\edef\@p@sheight{\@bbh}
				\edef\@p@swidth{\@bbw}
			\fi
		\fi
}
\def\compute@resv{
		\if@rheight \else \edef\@p@srheight{\@p@sheight} \fi
		\if@rwidth \else \edef\@p@srwidth{\@p@swidth} \fi
}
%
\def\compute@sizes{
	\compute@bb
	\compute@handw
	\compute@resv
}
%
%
\def\psfig#1{\vbox {
	%
	\ps@init@parms
	\parse@ps@parms{#1}
	\compute@sizes
	\ifnum\@p@scost<\@psdraft{
		\typeout{psfig: including \@p@sfile \space }
		\special{ps::[begin] 	\@p@swidth \space \@p@sheight \space
				\@p@sbbllx \space \@p@sbblly \space
				\@p@sbburx \space \@p@sbbury \space
				startTexFig \space }
		\if@clip{
			\typeout{(clip)}
			\special{ps:: \@p@sbbllx \space \@p@sbblly \space
				\@p@sbburx \space \@p@sbbury \space
				doclip \space }
		}\fi
		\if@prologfile
		    \special{ps: plotfile \@prologfileval \space } \fi
		\special{ps: plotfile \@p@sfile \space }
		\if@postlogfile
		    \special{ps: plotfile \@postlogfileval \space } \fi
		\special{ps::[end] endTexFig \space }
		\vbox to \@p@srheight true sp{
			\hbox to \@p@srwidth true sp{
				\hfil
			}
		\vfil
		}
	}\else{
		\vbox to \@p@srheight true sp{
		\vss
			\hbox to \@p@srwidth true sp{
				\hss
				\@p@sfile
				\hss
			}
		\vss
		}
	}\fi
}}
\catcode`\@=12\relax

\pageno=1                                      
\def\msun{M_{\odot}}
\def\kms{{\rm km \ s^{-1}}}
\def\au{{\rm AU}}
\def\pc{{\rm pc}}
\def\yr{{\rm yr}}
\def\myr{\msun \ \yr^{-1}}
\def\cc{{\rm cm^{-3}}}
\def\runningtitletext{DISK WINDS}
\def\runningauthortext{A. K\"onigl and R.E. Pudritz}

\null
\firstpageskip

{\baselineskip=14pt
\title{DISK WINDS AND }
\title{THE ACCRETION--OUTFLOW CONNECTION}
}

\vskip .3truein
\name{ARIEH K\"ONIGL}
\affiliation{Department of Astronomy \& Astrophysics, University of Chicago}
\affiliation{Chicago, IL 60637, USA}
\vskip .1truein
\leftline{and}
\vskip .1truein
\name{RALPH E. PUDRITZ}
\affiliation{Department of Physics and Astronomy, McMaster University}
\affiliation{Hamilton, Ontario L8S 4M1, Canada}
\vskip .3truein

\abstract{We review recent observational and theoretical results on
the relationship between circumstellar accretion disks and jets
in young stellar objects. We then present a theoretical framework
that interprets jets as accretion-powered, centrifugally driven 
winds from magnetized accretion disks.   
Recent progress in the numerical simulation of such outflows is described.
We also discuss the structure of the underlying
magnetized protostellar disks, emphasizing the role that
large-scale, open magnetic fields can play in angular momentum transport.}


\mainsection{I.~~INTRODUCTION}

Two of the most remarkable aspects of 
star formation are the presence of disks and
of energetic outflows already during the earliest phases of protostellar
evolution. There is now strong
evidence for an apparent correlation between the presence of
outflows and of actively accreting disks, which suggests that
there is a physical link between them. The prevalent
interpretation is that the outflows are powered by accretion and
that magnetic stresses mediate the inflow and outflow
processes and eject some of the inflowing matter from the disk
surfaces. If disks are threaded by open magnetic field lines
then the outflows can take the form of centrifugally driven
winds. Such highly collimated winds carry angular momentum 
and may, in principle,
play an important role in the angular momentum budget of 
disks and their central protostars. 

This review concentrates on the
developments in the study of outflows and their relationship to
circumstellar disks that have occurred since the publication
of {\it Protostars and Planets III}. The reader may consult 
K\"onigl and Ruden in that volume and 
Pudritz et al. (1991) for reviews of earlier work.
In Sec. II we summarize the observational findings on outflows, disks, and
magnetic fields, and their implications.
Sec. III deals with the general theory of magnetized
outflows and Sec. IV describes numerical simulations of
disk-driven MHD winds. In Sec. V we consider the
theory of magnetized protostellar disks.
Our conclusions are presented in Sec. VI.

\mainsection{{I}{I}.~~OBSERVATIONAL BACKGROUND}
\backup
\subsection{A.~~Bipolar Outflows and Jets}
Bipolar molecular outflows and narrow atomic jets are ubiquitous
phenomena in protostars. There are now more than 200 bipolar CO
sources known (see chapter by Richer et al.): they typically
appear as comparatively low-velocity ($\simlt 25 \ \kms$) and moderately
collimated (length-to-width ratios $\sim 3-10$) lobes, although
several highly collimated CO outflows that exhibit
high velocities ($>~40 \ \kms$) near the flow axis have now been detected.
The mass outflow rate exhibits a continuous
increase with the bolometric
luminosity of the driving source for $L_{\rm bol}$ in the range
$\sim 1-10^6 \, L_{\odot}$.
Molecular outflows are present through much of the
embedded phase of protostars and, in
fact, appear to be most powerful and best collimated during the
earliest (Class 0) protostellar evolutionary phase (Bontemps et
al. 1996).

The bipolar lobes are generally understood to
represent ambient molecular material that has been swept up by
the much faster, highly supersonic jets that
emanate from the central star/disk system (see chapters by
Eisl\"offel et al. and by Hartigan et al.). 
Jets associated with low-luminosity
($L_{\rm bol} < 10^3 \, L_{\odot}$) young stellar objects (YSOs)
have velocities in the range $\sim 150-400 \ \kms$, large ($>
20$) Mach numbers, and can have opening
angles as small as $\sim 3-5^{\circ}$ on scales of $10^3-10^4\
\au$. The inferred mass outflow rates are $\sim 10^{-10}-10^{-8}\
\myr$. A significant
number of outflows has also been detected by
optical observations of intermediate-mass ($2 \simlt 
M_*/M_{\odot} \simlt 10$) Herbig Ae/Be stars and
other high-luminosity sources (Mundt and Ray
1994; Corcoran and Ray 1997). The jet speeds and mass outflow
rates in these YSOs are, respectively, a factor $\sim 2-3$ and $\sim
10-100$ higher than in low-$L_{\rm bol}$ objects.
The total momentum
delivered by the jets, taking into account both the density
corrections implied by their partial ionization state and the long
lifetimes indicated by the detection of pc-scale outflows, appears to be
consistent with that measured in the associated CO outflows
(e.g., Hartigan et al. 1994; Eisl\"offel and Mundt 1997). A
critical review of the physical mechanisms of coupling the jets and
the surrounding gas is given in Cabrit et al. (1997).

The momentum discharge deduced from the bipolar-outflow observations
is typically a factor $\sim 10^2$ higher than the radiation-pressure
thrust $L_{\rm bol}/c$ produced by the central source (e.g.,
Lada 1985), which rules out radiative acceleration of the
jets. Since the bolometric luminosity of protostars
is by and large due to accretion, and since the ratio of jet
kinetic luminosity to thrust is of order the outflow speed
($\sim 10^{-3}\, c$), it follows that the jet kinetic luminosity
is on average a fraction $\sim 0.1$ of the rate at which gravitational
energy is liberated by accretion. This high ejection efficiency
is most naturally understood if the jets are driven magnetically.

Magnetic fields have also been implicated in jet
collimation. A particulary instructive case 
is provided by {\it HST} observations of the prototypical disk/jet
system HH 30 (Burrows et al. 1996).
The jet in this source can be traced to within $\simlt 30\ \au$
from the star and appears as a cone with an opening angle of 
3$^{\circ}$ between 70 and $700\ \au$. The narrowness of the
jet indicates some form of intrinsic collimation since external
density gradients would not act effectively on these small scales. Magnetic
collimation is a likely candidate, made even more plausible by
the fact that the jet appears to recollimate: its apparent opening
angle decreases to 1.9$^{\circ}$ between 350 and $10^4\ \au$.
Similar indications of recollimation have
also been found in other jets. Given that
any inertial confinement would be expected to diminish with 
distance from the source, this
points to the likely role of intrinsic magnetic collimation.

\subsection{B.~~Connection with Accretion Disks}

The evidence for disks around YSOs and for their link to
outflows has been strengthened by a variety of recent observations.
These include
systematic studies of the frequency of disks by means of
infrared and millimeter surveys and further interferometric
mappings (now comprising also the submillimeter range) that have resolved the
structure and velocity field of disks down to scales of a few
tens of AU (see chapter by Wilner and Lay). High-resolution
images of disks in several jet sources have also
been obtained in the near-infrared using adaptive optics and in
the optical using the {\it HST}\/ (see chapter by McCaughrean et al.).

Stellar jets are believed to be powered by the
gravitational energy liberated in the accretion process and to
be fed by disk material. This picture is supported by the
strong apparent correlation that is found (e.g., Cabrit et
al. 1990; Cabrit and Andr\'e 1991; Hartigan et al. 1995) between
the presence of outflow
signatures (such as P Cyg line profiles, forbidden line emission,
thermal radio radiation, or well-developed molecular lobes) and accretion 
diagnostics (such as ultraviolet, infrared, and millimeter-wavelength
emission excesses, or inverse P Cyg line profiles). Further
support is provided by the apparent decline in outflow activity
with stellar age, which follows the similar trend exhibited by
the disk frequency (see chapters by Andr\'e et al. and by Mundy
et al.) and mass accretion rate (see chapter by Calvet et al.). While
virtually every Class 0 source has an associated outflow,
a survey of optical and molecular outflows in the
Taurus-Auriga cloud (Gomez et al. 1997) found an incidence
rate of $\simgt 60\%$ among Class I objects but only $\sim 10\%$ among
Class II ones (and none in Class III objects). The
inference that jets are powered by
accretion and originate in accretion disks is strengthened by
the evidence for disks in the youngest YSOs in which outflows are detected.
For example, sub-mm interferometric observations of VLA 1623, one of the
youngest known Class 0 sources, imply the presence of a
circumstellar disk of radius $< 175\ \au$ and mass $\ge .03\,
\msun$ (Pudritz et al. 1996).

Corcoran and Ray (1998) demonstrated that the correlation
between [OI]$\lambda$6300 line luminosity (an outflow
signature) and  excess infrared luminosity (an accretion
diagnostic) originally found in Class II sources
extends smoothly to YSOs with masses of up to $\sim 10\, \msun$
and spans 5 orders of magnitude in luminosity.
It is noteworthy that correlations of the type $\dot
M \propto L_{\rm bol}^{0.6}$ that apply to both low-luminosity and
high-luminosity YSOs have been established in several
independent studies for the mass {\it accretion}\/ rate
(from IR continuum measurements; Hillenbrand et al. 1992 -- see,
however, Hartmann et al. 1993, Bell 1994, Miroshnichenko et
al. 1997, and Pezzuto et al. 1997 for alternative interpretations of
the infrared emission in Herbig Ae/Be stars), the
{\it ionized}\/ mass {\it outflow}\/ rate in the jets (from radio continuum
observations; Skinner et al. 1993), and the bipolar {\it
molecular outflow}\/ rate (from CO line measurements; Levreault
1988). Taken together, these relationships suggest that a strong
link between accretion and outflow exists also in high-mass YSOs
and that the underlying physical mechanism is basically the same
as in low-mass objects (see K\"onigl 1999).

Strong evidence for a disk origin of jets is available for the energetic
outflows associated with FU Orionis outbursts (see chapters by
Bell et al. and by Calvet et al.). The outbursts have
been inferred to arise in young YSOs that are still rapidly
accreting, although it is possible that they
last into the Class II phase. The duration of a typical outburst is $\sim
10^2\ \yr$, and during that time the mass accretion rate (as
inferred from the bolometric luminosity) is $\sim 10^{-4}\
\myr$, with the deduced mass outflow rate $\dot M_{\rm wind}$
(at least in the most powerful sources like FU Ori and Z CMa)
being a tenth as large. The ratio $\dot M_{\rm wind}/\dot M_{\rm
acc} \approx 0.1$ is similar to that inferred in Class II YSOs and
again points to a rather efficient outflow mechanism.
Detailed spectral
modeling demonstrates that virtually all the
emission during an outburst is produced in a rotating disk.
Furthermore, the correlation
found in the prototype FU Ori between the strength and
the velocity shift of various
photospheric absorption lines can be naturally interpreted in
terms of a wind accelerating from the disk surface (Calvet et
al. 1993; Hartmann and
Calvet 1995). Because of the comparatively low temperatures
($\sim 6000$ K) in the wind acceleration zone, thermal-pressure
and radiative driving are unimportant (the latter due
to the lack of a high-temperature source): magnetic driving is
thus strongly indicated. The recurrence time of outbursts has
been estimated to lie in the range $\sim 10^3-10^4\ \yr$, and if
these outbursts are associated with the large-scale bow shocks
detected in pc-scale jets (e.g., Reipurth 1991), then a value
near the lower end of the range is implied.  In that case most
of the stellar mass would be accumulated through this process,
and, correspondingly, most of the mass and momentum ejected over
the lifetime of the YSO would originate in a disk-driven outflow
during the outburst phases (Hartmann 1997).

\subsection{C.~~Magnetic Fields in Outflow Sources}

The commonly accepted
scenario for the origin of low-mass protostars is that they are produced
from the collapse of the inner regions of molecular clouds that
are supported by large-scale magnetic fields (and likely also hydromagnetic
waves). In this picture, a gravitationally unstable inner core
forms as a result of mass redistribution by ambipolar diffusion
and subsequently collapses dynamically (see review by {\nobreak McKee} et
al. 1993). The mass accretion rates predicted by this picture
are consistent with the inferred evolution of young YSOs (e.g.,
Ciolek and K\"onigl 1998).
Basic support for this scenario is provided by
far-infrared (e.g., Hildebrand et al. 1995) and sub-mm (e.g., Greaves et
al. 1994, 1995; Schleuning 1998; Greaves and Holland 1998)
polarization measurements, that reveal an ordered,
hourglass-shaped field morphology on sub-pc scales consistent
with the field lines
being pulled in at the equatorial plane of the contracting core. 
Moreover, ${\rm H\ I}$ and OH Zeeman measurements (e.g., Crutcher
et al. 1993, 1994, 1996) are consistent with the magnetic
field having the strength to support the bulk of the cloud
against gravitational collapse.

There now exist measurements of magnetic fields in the flows
themselves at large distances
from the origin (see chapter by Eisl\"offel et al.). In
particular, the strong circular polarization detected in T Tau S in two
oppositely directed nonthermal emission knots separated by $20\
\au$ indicates a field strength of at least several gauss (Ray
et al. 1997). This high value can be attributed to a
magnetic field that is advected from the origin by the associated
stellar outflow and that dominates the interal energy of the
jet. This observation thus provides direct evidence for the essentially
hydromagnetic character of jets. 

\mainsection{{I}{I}{I}.~~ MHD WINDS FROM ACCRETION DISKS}
\backup
\subsection{A.~~Basic MHD Wind Theory}
The theory of centrifugally driven winds was first formulated
in the context of rotating, magnetized stars
(Schatzmann 1962; Weber and Davis 1967; Mestel 1968). Using
1-D, axisymmetric models, it was shown that such stars could
lose angular momentum by driving winds of this type.
This idea was applied to magnetized accretion disks 
in the seminal paper of Blandford and Payne (1982, BP).
Every annulus of a Keplerian disk
may be regarded as rotating close to its ``breakup'' speed, 
so disks are ideal drivers of outflow when sufficiently 
well magnetized. 
The removal of disk angular momentum allows matter to move inward and
produces an accretion flow.
In a steady state, field lines must also slip radially out of the 
accreting gas and maintain
their fixed position in space: this necessitates
diffusive processes.
As we discuss in Sec. V.B, strong field diffusivity is a natural
attribute of the partially ionized regions of protostellar
disks, and it could counter both the advection of the
field lines by the radial inflow and their winding-up by
the differential rotation in the disk.

Consider, for the moment, the simplest possible description of a 
magnetized,
rotating gas threaded by a large-scale, open field
(characterized by an even symmetry about the midplane $z=0$).
The equations of stationary, axisymmetric, ideal MHD are the conservation  
of mass (continuity equation); 
the equation of motion with conducting gas of
density $\rho$ subject to forces associated with the
pressure $P$, the gravitational field (from the central
object whose gravitational potential is $\phi$), 
and the magnetic field ${\bf B}$; the induction equation   
for the evolution of the magnetic field in the moving
fluid; and the 
solenoidal condition on ${\bf B}$:

$$ \nabla \cdot ( \rho {\bf V}) = 0\; ,  \eqno(1)$$

$$ \rho {\bf V \cdot \nabla V} = - \nabla p - \rho \nabla \phi
           + { 1 \over 4 \pi} ( {\bf \nabla \times B}) \times
	   {\bf B}\; ,
\eqno(2)
$$

$$ {\bf \nabla \times (V \times B) } = 0 \; ,
\eqno(3) $$ 

$$ {\bf \nabla \cdot B} = 0 \; . \eqno(4)  $$
\noindent

Consider the {\bf angular momentum equation} for axisymmetric
flows.   This is described by the  $ \phi $ component
of equation (2).  Ignoring stresses that would arise due to 
turbulence, and noting that neither the
pressure nor the gravitational term contributes, one
finds that 

$$ \rho {\bf V_{\rm p}} \cdot \nabla (r V_{\phi}) = { {\bf B_{\rm p}} \over
4 \pi} \cdot \nabla (r B_{\phi})\; ,   \eqno(5)   $$

\noindent
where we have broken the magnetic and velocity fields
into poloidal and toroidal components:
$  {\bf B =  B_{\rm p}} +  B_{\phi}
{\bf \hat e_{\phi}} $ and 
$ {\bf V = V_{\rm p}} + V_{\phi} {\bf \hat e_{\phi}}$.       

Important links between the velocity field and the magnetic
field are contained in 
the induction equation (3), whose solution is
$${\bf V \times
B } = \nabla \psi,  
\eqno(6)
$$  
where $\psi$ is some scalar potential.  This shows that 
the electric field due to the bulk motion of conducting
gas in the magnetic field is derivable from an electrostatic
potential. This has two important
ramifications.  The first is that, because of axisymmetry, 
the toroidal component of this equation must vanish
($\partial \psi / \partial \phi = 0$).  This  
forces the poloidal velocity vector to be parallel to
the poloidal component of the magnetic field, ${\bf V_{\rm p} \vert \vert B_{\rm p}}$.  
This, in turn, implies that there is a function 
$ k$, the mass load of the wind, such that
$$
\rho {\bf V_{\rm p}} = k {\bf B_{\rm p}}\; .
\eqno(7)
$$
\noindent
Substitution of this result into the continuity equation (1), and
then use of the solenoidal condition
(eq. [4]) reveals
that $k$ is a constant along a surface of constant magnetic
flux, i.e., that it is conserved along field lines.
This function can be more revealingly cast by noting
that the wind mass loss rate passing through an 
annular section of the flow 
of area $dA$ through the flow is $d \dot M_w = \rho V_{\rm p} dA$,
while the amount of poloidal magnetic flux through this
same annulus is $d \Phi = B_{\rm p} dA $.
Thus, the mass load per unit time and per unit magnetic flux,
which is preserved along each
streamline emanating from the rotor (a disk
in this case), is
$$
k = { \rho V_{\rm p} \over B_{\rm p}} = {d \dot M_w \over d \Phi} \; . \eqno(8)
$$ 
The mass load is determined by the physics of the 
underlying rotor, which is its source.  

A second major consequence of the induction equation follows
from the poloidal part of equation (6).  Taking the 
dot product of it with ${\bf B_{\rm p} }$ and using equation (7),
one easily proves that
$ \psi$ is also a constant along a magnetic flux surface
and that it must take the form
$ \psi = \Omega - (k B_{\phi} / \rho r)$.  
In order to evaluate $\psi$, note that $B_{\phi} = 0$ 
at the disk midplane by the assumed even
symmetry. Thus $\psi$ equals $\Omega_0$, the angular
velocity of the disk at the midplane.  One thus has a
relation between the toroidal field in a rotating flow
and the rotation of that flow,
$$
B_{\phi} = {\rho r \over k} ( \Omega - \Omega_0)\; .
\eqno(9)
$$
 
Let us now examine the angular momentum equation.
Returning to the full equation (5) and applying
equation (7) and the constancy of $k$ 
along a field line, one obtains
$$ {\bf B_{\rm p} \cdot \nabla}  (r V_{\phi} - {r B_{\phi} \over 4 \pi k}
) = 0\; .
\eqno(10)
$$
\noindent
Hence the angular momentum per unit mass,
$$ l = r V_{\phi} - {r B_{\phi} \over 4 \pi k}\; , \eqno(11)
$$
is constant along a streamline.
This shows that the specific angular momentum of a magnetized flow
is carried by both the rotating gas (first term) and
the twisted field (second term).  
The value of $l$
may be found by eliminating the toroidal field between equations
(9) and (11) and solving for the rotation speed
of the flow,
$$
r V_{\phi} = { lm^2 -  r^2 \Omega_o^2 \over m^2 - 1}\; ,
\eqno(12)
$$
where the Alfv\'en Mach number $m$ of the flow
is defined as $m^2 = V_{\rm p}^2 / V_{\rm A}^2$, with $V_{\rm A} = B_{\rm p} /( 4 \pi \rho)^{1/2}$
being the Alfv\'en speed of the flow.  The Alfv\'en surface
is the locus of the points $r = r_{\rm A}$ on the outflow field lines
where $m=1$.  The flow along
any field line essentially corotates 
with the rotor until
this point is reached. 
>From the regularity condition at the
Alfv\'en critical point (where the denominator of eq. [12] vanishes)
one infers that the conserved specific angular momentum
satisfies
$$
l = \Omega_0 r_{\rm A}^2 \; .
\eqno(13)   
$$
If we imagine following a field line from its 
footpoint at a radius $r_0$, the Alfv\'en radius is at a distance  
$r_{\rm A}(r_0)$ from the rotation axis
and constitutes the lever arm
for the
back torque that this flow exerts on the disk.
The other 
critical points of the  outflow are where the outflow speed $V_{\rm p}$ equals
the speed of the slow and fast magnetosonic modes
in the flow (at the so-called SM and FM surfaces).  

Finally, a generalized version of Bernoulli's equation
may be derived by taking the dot product of the
equation of motion with ${\bf B_{\rm p}} $.  One then finds
that the specific energy
$$
E = {1 \over 2} ( V_{\rm p}^2 + \Omega^2 r^2) + \phi + h +
         \Omega_0(\Omega_0 r_{\rm A}^2 - \Omega r^2)\; ,
\eqno(14)
$$
where $h$ is the enthalpy per unit mass, is also a field-line
constant.

The terminal speed $V_{\rm p} = V_{\infty}$ corresponds to the
region where the
gravitational potential and the rotational energy of the flow
are negligible.  Since for cold flows the specific enthalpy may
be ignored, one infers from equation (14)
$$
V_{\infty} \simeq 2^{1/2} \Omega_0 r_{\rm A} \; ,
\eqno(15)
$$ 
a result first obtained by Michel (1969) for 1-D
flows.     
The important point regarding outflow speeds from disks
is that $V_{\infty} / \Omega_0 r_0 \approx 2^{1/2} r_{\rm A} / r_0$:
the asymptotic speed is larger than the rotor speed
by a factor that is approximately the ratio of the lever arm to
the footpoint radius.

\subsection{B.~~Connection with Underlying Accretion Disk}
We now apply the angular momentum conservation relation (5) to calculate
the torque exerted on a thin accretion disk 
by the external magnetic field.  The vertical flow speed in the 
disk is negligible, so only the radial inflow  speed $V_{\rm r}$ 
and the rotation speed $V_{\phi}$ (Keplerian for thin disks) contribute. 
On the right-hand side, both the radial and vertical magnetic
contributions come into play, so

$$ {\rho V_r \over r_0} { \partial (r_0 V_{\phi}) \over \partial r_0} = 
                {B_{\rm r} \over 4\pi r_0 } {\partial (r_0 B_{\phi}) \over
                     \partial r_0} + {B_{\rm z} \over
		     4\pi}{\partial B_{\phi} \over \partial z} \; . 
\eqno(16)
$$   
\noindent
One sees that specific angular momentum is removed from
the inward accretion flow by the action of
two types of magnetic torque.  The first term on the 
r.h.s. represents radial angular momentum associated with the radial
shear of the toroidal field, while the second term is
vertical tranport due to the vertical shear of the 
toroidal field.  In a thin disk,
and for typical field inclinations,
the second term
will dominate.  Note that the first
term vanishes at the disk midplane because $B_{\rm r} = 0$ there.
Now, following standard thin-disk theory, vertical integration 
of the resulting equation gives a relation between the disk accretion
rate, $\dot M_{\rm acc} = - 2 \pi \Sigma V_r r_0$, and the 
magnetic torques acting on its surfaces (subscript s),
$$
\dot M_{\rm acc} { d (r_0 V_{\phi}) \over dr_0} = - r_0^2 B_{\rm \phi,s}
B_{\rm z}\; ,
\eqno(17)
$$ 
Angular momentum is thus extracted
out of disks threaded by open magnetic fields.  The
angular momentum can be
carried away either by toroisional Alfv\'en waves or,
when the magnetic field lines are inclined by more than $30^{\circ}$ from
the vertical, by a centrifugally driven wind.
By rewriting equation (11) as $rB_{\phi} = 4\pi k (r V_{\phi} - l)$  
and using the derived relations for $k$ and $l$,
the disk angular momentum equation can be cast into its most fundamental form,
$$
\dot M_{\rm acc} {d( \Omega_0 r_0^2) \over dr_0} = {d \dot M_{\rm wind} \over d r_0}
                   \Omega_0 r_{\rm A}^2
                   (1 - (r_0/r_{\rm A})^2)\; .
\eqno(18)
$$               
This equation shows that there is a crucial link between the 
mass outflow in the wind and the mass accretion rate 
through the disk:
$$
\dot M_{\rm acc} \simeq (r_{\rm A}/ r_0)^2 \, \dot M_{\rm wind}\; .
\eqno(19)
$$
One has arrived at the 
profoundly useful expression of the 
idea that,
if viscous torques in the disk are relatively unimportant,
the rate at which the disk
loses angular momentum ($\dot j_d = \dot M_{\rm acc} \Omega_0 r_0^2$)
is exactly the rate at which it is carried away by the wind
($\dot j_w = \dot M_{\rm wind} \Omega_0 r_{\rm A}^2$).
 
The value of the ratio $r_{\rm A} / r_0$ is $\sim 3$
for typical parameters, so one finds $\dot M_{\rm wind} / \dot
M_{\rm acc} \simeq 0.1$, which is in excellent agreement
with the observations (Sec. II).
The explanation of this relationship is thus
intimately linked to the disk's angular momentum loss to the wind.  

\subsection{C.~~Flow Initiation and Collimation}
Stellar winds usually require hot coronae to get started, while 
winds from disks, which effectively rotate near ``breakup,'' do
not.  In the case of a thin disk even a cool atmosphere will suffice as 
long as the field lines
emerging from the disk make an angle 
$\le 60^{\circ}$ to the surface. This follows from Bernoulli's
equation by comparing the variations in the
effective gravitational potential and the kinetic energy of a
particle that moves along a field line near the disk surface
(BP; K\"onigl and Ruden 1993; Spruit 1996; but see Ogilvie and Livio 1998).

The collimation of an outflow, as it accelerates away
from the disk, arises
to a large extent
from the hoop stress of the toroidal field component.  From equation (9) one
sees that at the Alfv\'en surface
$|B_{\phi}| \simeq B_{\rm p}$ and that in the far field
(assuming that the flow opens up to radii $r >> r_{\rm A}$)
$B_{\phi} / B_{\rm p} \simeq r/ r_{\rm A}$.  Thus, the inertia
of the gas, forced to corotate with the outflow out
to $r_{\rm A}$, causes the jet to eventually
self-collimate through the ${\bf j_z \times B_{\phi}}$
force, which is known as the    
z-pinch in the plasma physics literature.

The detailed radial structure of the outflow
is deduced by balancing all forces perpendicular
to the field lines and is described by   
the so-called Grad-Shafranov equation.   
This is a complicated nonlinear equation for which
no general solutions are available.
Because of the mathematical difficulties 
(e.g., Heinemann and Olbert 1978),  
the analytic studies have been characterized by simplified
approaches, including separation of variables
(e.g., Tsinganos and Trussoni 1990; Sauty and Tsinganos 1994),
self-similarity (e.g., BP; Bacciotti and Chiuderi 1992;
Contopoulos and
Lovelace 1994; Lynden-Bell and Boily 1994),
previously ``guessed'' magnetic
configurations (e.g., Pudritz and Norman 1983; Lery et al. 1998), as well
as examinations of various asymptotic limits to the theory
(e.g., Heyvaerts and Norman 1989, HN; Appl and Camenzind 1993;
Ostriker 1997). Pelletier and Pudritz (1992)
constructed non--self-similar models of
disk winds including cases where the wind emerges only from a
finite portion of the disk.

Self-similarity 
imposes a specific structure on the underlying disk. In the BP
model, all quantities scale as power laws of spherical radius
along a given direction. This directly implies that
the Alfv\'en surface is conical and, similarly, that the
disk's scale height $H(r)$ scales linearly with $r$.
Furthermore, inasmuch as the problem contains a characteristic speed, namely
the Kepler speed in the disk, one infers that
$V_{\rm A} \propto C_{\rm s} \propto V_{\rm r} \propto V_{\infty}
\propto V_{\rm K}$, i.e., that the  Alfv\'en, sound, radial inflow,
terminal outflow, and Kepler speeds, respectively, are all proportional
to one another.  The scaling $C_{\rm s} \propto V_{\rm K}$
implies that the disk temperature has the virial 
scaling $T \propto r^{-1}$.  The scaling $V_{\rm r} \propto
V_{\rm K}$ as
well as the $H(r)$ relation imply that,
for a constant mass accretion rate
$\dot M_{\rm acc} = - 2 \pi ( 2 H \rho)
V_{\rm r} r$, the density $\rho \propto r^{-3/2}$.  Next, the scaling of
the disk Alfv\'en speed $V_{\rm A} \propto V_{\rm K}$ together with
the density result imply that the disk poloidal field (and
hence also $B_\phi$) scales as $r^{-5/4}$. In turn,
the mass load $k_0 \propto r_0^{-3/4}$ and the wind mass loss rate 
$\dot M_{\rm wind} (r_0) \propto ln \, r_0$.
More general self-similar models may be constructed by
making the more realistic assumption that  
mass is lost from the disk so that the accretion rate is not constant; 
$\dot M_{\rm acc}(r) \propto r^{-\mu}$.  
Adopting the radial scaling of the disk magnetic
field, $B_0(r) \propto r^{-\nu}$, one finds 
that self-similarity imposes the scaling
$\mu = 2(\nu - 1.25)$ (see eq. [17]). 

The work of HN and others (see Heyvaerts and Norman, 1997)  
has shown that in general two types of collimation are possible, depending on
the asymptotic behavior of the current intensity $I = (c/2)
r B_{\phi} \propto r^2 \rho \Omega_0$.  If $I \rightarrow 0$ as
$r \rightarrow \infty $, then the field lines are space-filling
paraboloids, whereas if this limit for the current
is finite, then the flow is collimated to cylinders.   
The character of the flow therefore depends on
the boundary conditions at the disk. 

Finally, we note that
the stability of magnetized jets with toroidal magnetic fields
is strongly assisted by the jet's poloidal magnetic field, which
acts as a spinal column for the jet (e.g., Appl and Camenzind 1992). 

\mainsection{{I}{V}.~~NUMERICAL SIMULATIONS OF DISK WINDS}

The advent of numerical simulations has  
finally made it possible to study the rich, 
time-dependent behavior of MHD disk winds.
This allows one to test the stationary theory
presented above as well as to search for the conditions
that give rise to
episodic outflows.
These simulations represent perhaps the main advance in the
subject since {\it Protostars and Planets III}.
The published 
simulations generally assume ideal MHD and 
may be grouped into two classes: 1) {\it dynamic 
MHD disks}, in which the structure and evolution of 
the magnetized disk is also part of the simulation, and
2) {\it stationary MHD disks}, in which the underlying accretion disk
does not change and provides fixed boundary conditions for
the outflow problem.  

\subsection{A.~~Dynamic Disks and Winds}  
The first numerical calculations of disk winds were 
published by Uchida and Shibata (1985, US), and
Shibata and Uchida (1986).  These simulations modeled 
a magnetized disk in sub-Keplerian rotation  
and showed that a 
rapid radial collapse develops in which the initially poloidal 
field threading the disk is wound up due to the differential
rotation.  The vertical Alfv\'en speed, being smaller than
the free-fall speed, implies that a strong, vertical
toroidal field pressure gradient 
$\partial B_{\phi}^2 / \partial z$
must rapidly build up.   
This force results in the transient ejection
of coronal material above and below the disk 
as the spring uncoils.  The work of US was confirmed by
Stone and Norman (1994, SN) using their ideal-MHD, ZEUS 2-D
code (Stone and Norman 1992).

If the mean magnetic
field energy density is less than the thermal pressure in the disk 
($V_{\rm A} < C_{\rm s}$), 
then a strong magnetorotational instability will develop (e.g.,
Balbus and Hawley 1991, BH).
In 2-D, this leads to 
a vigorous, radial, channel flow and rapid outward transport of  
angular momentum. 
SN ran a series of ZEUS 2-D simulations for a uniform
magnetic field threading a wedge-shaped disk and a surrounding
corona (models defined by 4 parameters). 
Their simulations investigated three cases
(see also Bell and Lucek 1995 and Matsumoto et al. 1996): (A) sub-Keplerian
rotation, 
(B) a Kepler disk with strong
disk field,  
and (C) a Kepler disk with a
weak field.  
In Case A, rapid
collapse immediately ensues with an expanding, transient
outflow appearing in 2.5 orbits (reproducing US).
In Case B, the disk is BH-stable, but collapse occurs
anyway because of the very strong braking 
of the disk due to the external MHD torque.
In case C one again
sees rapid radial collapse of the disk, this
time because of a strong BH instability.

The 2-D channel flow does not, however, persist in 3-D: in
that case the BH instability develops into a fully turbulent
flow and the inflow rate is significantly reduced (see chapter
by Stone et al.).

\subsection{B.~~Stationary Disks and Winds}
The outflow problem can be clarified by  
focusing on the more restricted question of how a wind 
is accelerated and collimated for a prescribed set of fixed 
boundary conditions on the disk.  Keplerian disks in 
3-D are stable on many tens of orbital times and this  
justifies the stationary disk approach:
the launch and collimation of jets 
from their surfaces occurs in only a few inner-disk rotation times.  
Groups that have taken
this route include
Ustyugova et al. (1995), Ouyed et al. (1997, OPS),
Ouyed and Pudritz (1997a,b, OPI and OPII), Romanova et al. (1997),
and Meier et al. (1997).  The published simulations differ
in their assumed initial conditions, such as the
magnetic field distribution on the disk, 
the plasma $\beta$ ($ \equiv P_{\rm gas} / P_{\rm mag}$) 
above the disk surfaces, the state of the initial
disk corona, and the handling of
the gravity of the central star.  
Broadly speaking, all of the existing
calculations show that winds from accretion disks
can indeed be launched and accelerated, much  
along the lines suggested by the theory presented
in Sec. III.   The results differ, however, in the degree
to which flow collimation occurs.

Ustyugova et al. (1995) and Romanova et al.
(1997) employed a magnetic configuration
described by a monopole field centered beneath
the disk surface.
Nonequilibrium initial conditions 
as well as a softened gravitational potential were used.  
Relatively low resolution simulations of flows 
with $\beta >> 1$ (Ustyugova et al.)
showed that collimated, non-stationary outflows develop. 
Similar simulations in the strong-field ($\beta <1$) regime  
(Romanova et al.)
resulted in  
stationary, but uncollimated outflows on these scales. 

Simulations by OPS, OPI, and OPII 
(see Pudritz and Ouyed 1997 for a review) 
employed the 
ZEUS 2-D code
and studied two different magnetic configurations.
The coronal gas was initially taken to be in hydrostatic
equilibrium with
the central object as well
as in pressure balance with the top of the disk.
Two initial magnetic configurations were adopted, each
chosen so that no 
Lorentz force
is exerted on the 
hydrostatic corona, 
i.e., such that ${\bf j} = 0$.  
These are the potential field configuration of
Cao and Spruit (1994), 
and a uniform, vertical field that is
everywhere parallel to the disk rotation axis. 
The gravity was unsoftened in these calculations, 
$ (500 \times 200)$ spatial zones were used, and
simultations were run up to $400 t_{\rm i}$   
(where $t_{\rm i}$ is the Kepler time for an orbit
at the inner edge of the disk, $r_i$). 
This model is described by 5 parameters set at $r_{\rm i}$:
3 to describe
the initial corona (e.g., $\beta_{\rm i} = 1.0$), 
as well
as 2 parameters to describe the disk physics
(e.g., the injection speed $V_{\rm inj}$ of 
the material from the disk into the base of the
corona). 
The initial conditions correspond to turning
on the rotation of the underlying Keplerian disk at $t=0$.   

The first thing that happens
is the launch and propagation of a brief 
transient, torsional 
Alfv\'en wave front that sweeps out from the disk surface
but that leaves the corona largely undisturbed. 
An outflow also 
begins almost immediately and develops into
a stationary or an episodic jet.  These jet-like outflows 
are highly collimated and 
terminate in a jet shock.   
A bow shock, driven by the jet, pushes through the
corona.  A noticeably empty cavity dominates most of the volume
behind the bow shock: the cavity is filled by the toroidal
magnetic field that is generated by the jet itself.     
This may provide an explanation 
of the extended bipolar cavities
that surround highly collimated jets.
For a fiducial injection speed of $10^{-3} V_{\rm K}$, 
the potential field configuration
developed into a stationary outflow with many
(but not all, since the flow is not self-similar) of the
characteristics of the BP 
solution. 
The Alfv\'en surface
is correctly predicted by the analysis in  
section III. The Alfv\'en and fast-magnetosonic Mach numbers reach 
5 and 1.6, respectively, at $z = 10r_{\rm i}$, with the toroidal-to-poloidal
field-strength ratio being $\sim 3$ on this scale.
Outflow only takes place on field
lines that are inclined by less than $60^{\circ}$ with respect
to the disk surface, as predicted by BP. 

Figure 1 (adapted from OPS) shows that the poloidal
field lines in the initial state are collimated toward 
the rotation axis by the hoop stress of the jet's
toroidal field. The figure also shows
the propagation of the jet-driven bow-shock
and the eventual creation of a cylindrically collimated
jet with well determined
Alfv\'en and fast-magnetosonic critical surfaces
in the acceleration region above the disk 
(the slow-magnetosonic surface is too close to the disk to be
resolved in this figure).

{\psfig{figure=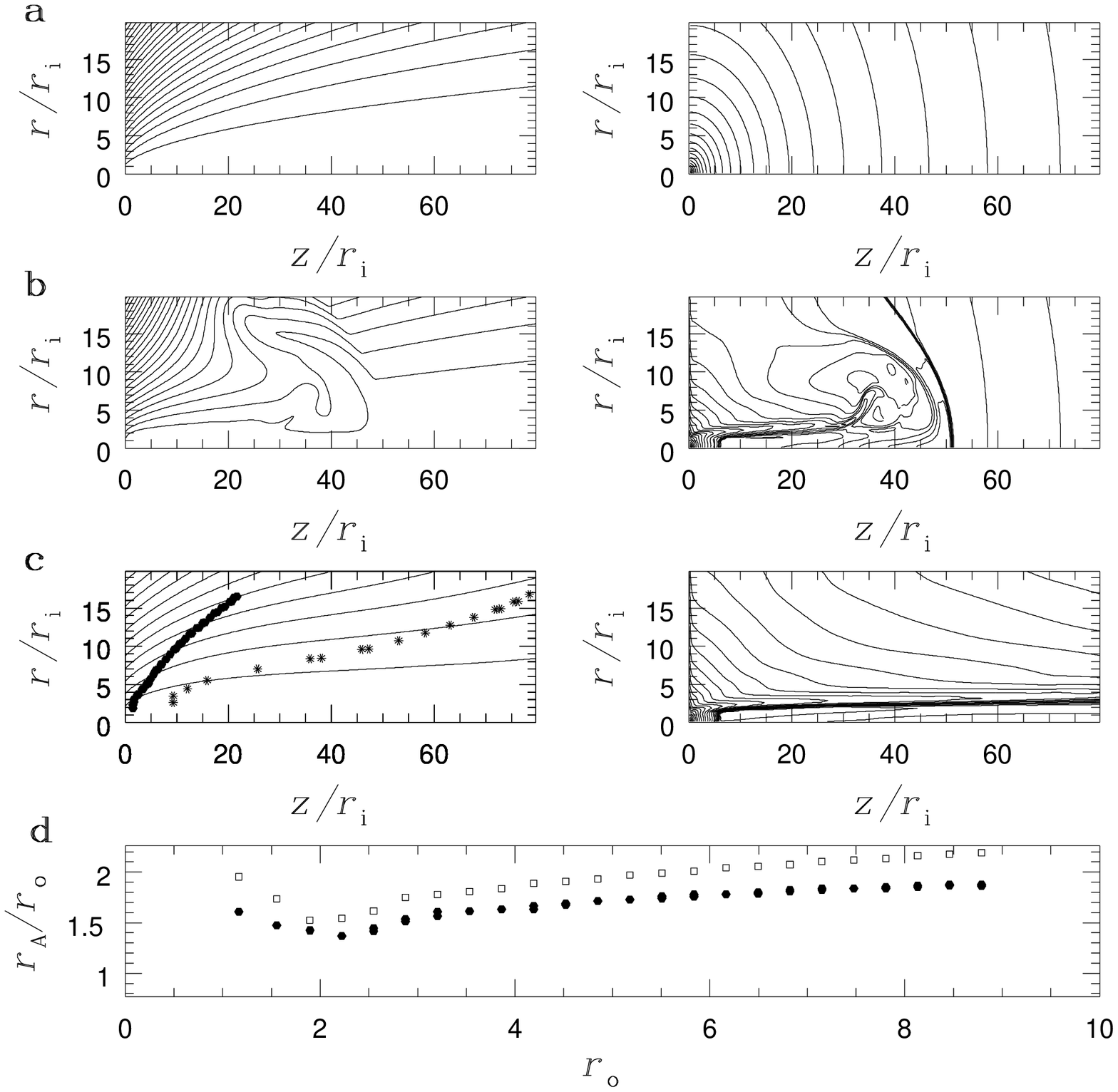,height=3.7in,width=4.2in} 
\caption{Figure 1.\capskip Numerical simulations of disk-driven
MHD outflows (adapted from OPS). In frame $a$, the left panel shows the
initial magnetic configuration corresponding to the ``potential
field'' solution, whereas the right panel displays the initial
isodensity contours of the corona. The flow injection speed is
$V_{\rm inj}=10^{-3} V_{\rm K}$. Frames $b$ and $c$ show the evolution
of the initial magnetic and density structures (the left and
right panels, respectively)at $100$ and $400$ inner time units. Frame
$c$ also displays the locations of the Alfv\'en critical surface
(filled hexagons) and the FM surface (stars).
In frame $d$, the Alfv\'en lever arm ($r_{\rm A}/r_{0}$) found
in the simulation is shown (filled hexagons) as a function of
the location ($r_0$) of the footpoints of the field
lines on the disk, and is compared to the prediction from the
steady-state theory of Sec. III.A (squares). 
}}

\bigskip
For the same boundary and initial conditions, the 
initially vertical field configuration leads to the development
of a jet that is episodic over the 400 $t_{\rm i}$ duration of the 
simulation (OPS, OPII).   Even though the initial configuration
is highly unfavorable to jet formation,
jet production occurs.  The reason for this is
the effect of the toroidal
field in the corona, which is concentrated towards the
inner edge of the disk (where the Kepler rotation
is the greatest).  It therefore exerts a radial pressure force
$\partial (B_{\phi}^2/8 \pi) / \partial r$ that 
opens up
field lines at larger radii.  As long as $\beta \simeq 1$ in this
region, field lines are pliable enough to move,
and a jet is launched.  Episodic knots are produced on a time
scale $\tau_{knot} \simeq r_{jet}/ V_{A,\phi}$, which
is reminiscent of a type of kink instability.  Regions
of high toroidal field strength and low density separate
and confine the knots, which in turn have high density and low
toroidal field strength.

The appearance of stationary or episodic
jets has nothing to do with the
magnetic configuration, but rather 
with
the mass loading of the magnetic configuration, as is seen in an extensive
set of simulations (Ouyed and Pudritz 1998).  
Figure 2 shows a simulation with 
the same potential configuration parameters as in Figure 1
except for a reduced injection speed (and hence mass load) of  
$V_{\rm inj} = 10^{-5} V_{\rm K}$.    
Clear episodic behavior is now seen in this flow.

{\psfig{figure=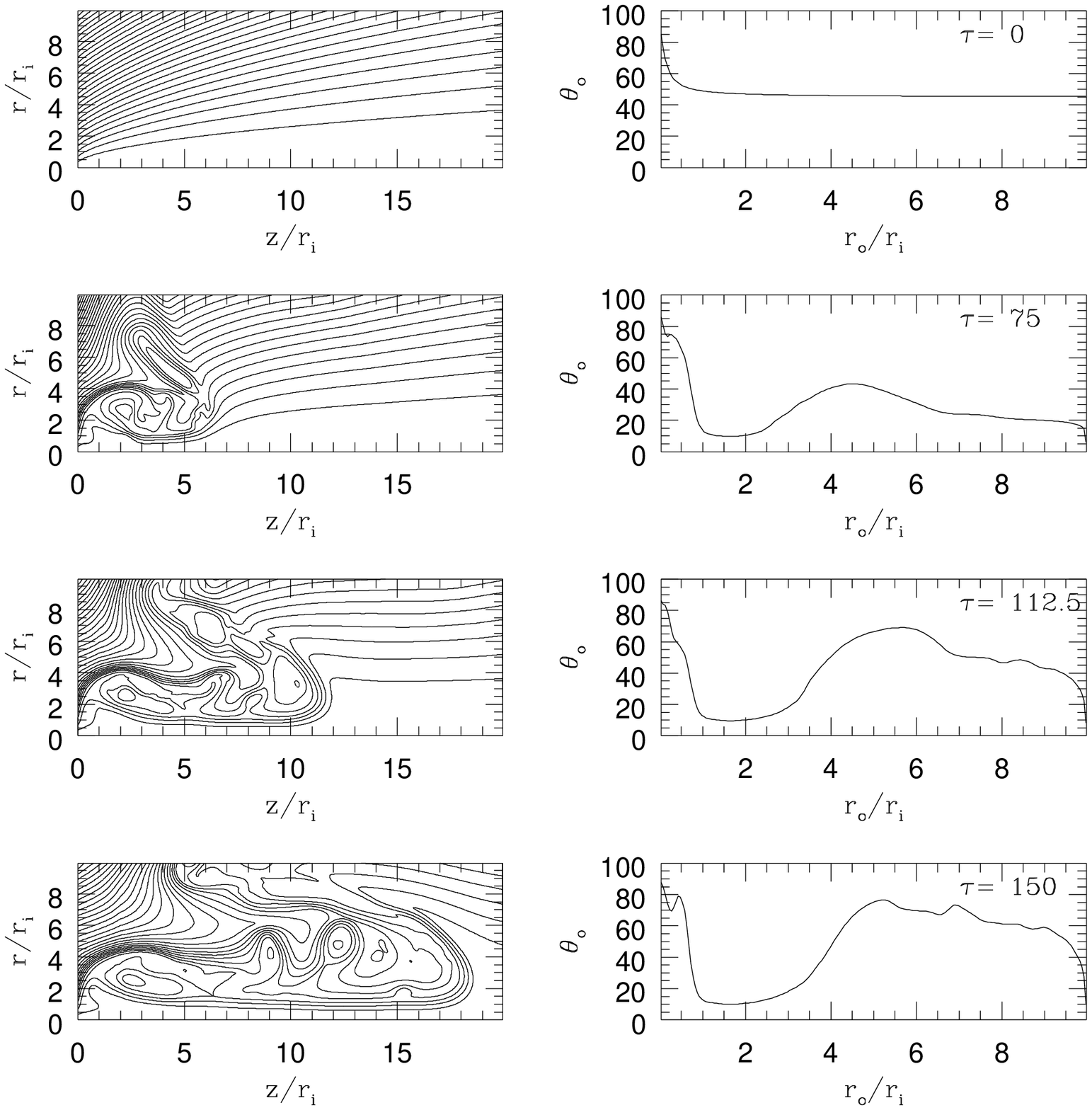,height=3.7in,width=4.2in} 
\caption{Figure 2.\capskip MHD outflow simulations with the same
parameters and magnetic configuration as in Fig. 1, except
$V_{\rm inj}=10^{-5} V_{\rm K}$ (adapted from Ouyed
and Pudritz 1998). The left panels show the
magnetic field structure and knot formation
at times 0.0, 75.0, 112.5, and 150.0 $t_{\rm i}$.
The right panels show the angle $\theta_0$ between the field lines and
the disk surface at these 4 times. Only
field lines that open up to an angle $\theta_0 \le 60^{\circ}$
are seen to drive an outflow.
}}

\bigskip
Nonsteady outflows could conceivably also arise in
highly conducting disk regions where the winding-up of the field
lines might eject gas through a strong $|B_{\phi}|$
pressure gradient (e.g., Contopoulos 1995). 
 
\subsection{C.~~Far-Field Behavior of MHD Jets}
Simulations of MHD jets propagating
into a uniform medium show that, in certain
respects, they differ dramatically 
from hydrodynamic jets.  
Clarke et al. (1986), Lind et al. (1986),
and K\"ossel et al
(1990) showed that the strong toroidal field of 
a jet prevents matter from spraying sideways
on encountering the jet shock.  Rather, jet material 
decelerated by a Mach disk and a strong annular
shock is focused mainly forward into a nose cone. This 
contrasts with the backflowing cocoon that characterizes
purely hydrodynamic, 
low-density jets.
Hydromagnetic jets
are generally preceeded by stronger, more oblique shocks and 
advance far more quickly into the surrounding medium
than their hydrodynamic counterparts 
(see K\"ossel et al. 1990 for more detail).
MHD jets may thus explain 
why one does not measure strong 
transverse expansions in bipolar molecular lobes 
(e.g., Masson and Chernin 1993; Cabrit 1997).

\mainsection{{V}.~~MAGNETIZED PROTOSTELLAR ACCRETION DISKS}
\backup
\subsection{A.~~Magnetic Angular Momentum Transport}
One of the primary reasons why open, ordered magnetic fields are
important in accreting astrophysical systems is that they can
mediate a vertical transport of angular momentum from
accretion disks. In Sec. III we discussed the angular momentum
transport by centrifugally driven winds, but it is important to
realize that angular momentum can be removed from the disk surfaces
even if the field-line inclination with respect to the vertical
is not large enough for the wind launching condition to be
satisfied, so long as the field is attached to a ``load'' that
exerts a back torque on the disk. In that case the disk can lose
angular momentum through torsional Alfv\'en waves propagating
away from the disk surfaces
(the ``magnetic braking'' mechanism; see
Mouschovias 1991 for a review). When the magnetic fields are
well below equipartition with the gas pressure in a
differentially rotating disk, the magnetorotational (Balbus-Hawley)
instability develops on a dynamical time scale and
produces a magnetic stress-dominated turbulence that gives rise to
radial transport of angular momentum characterized by
an effective viscosity parameter $\alpha$ that lies in the range
$\sim 0.005-0.5$ (see chapter by Stone et al.). 

Numerical simulations (see Sec. IV.A) have
demonstrated that all of the above processes could contribute to
the angular momentum transport in magnetized disks. One can
estimate the relative roles of vertical wind transport and radial
turbulent transport by taking the ratio of the 
external wind torque computed in Sec. III and the viscous 
torque associated with magnetocentrifugal turbulence, which can
be written as $ (B_{\rm z}^2 / 4 \pi P_{\rm gas})\, (r_{\rm A}/ \alpha H)$
(Pelletier and Pudritz 1992). Thus, even
if the ordered field threading the disk were far below
equipartition with the disk gas pressure,
a wind torque could still dominate a turbulent torque
simply because of its large lever arm (the Alfv\'en radius
$r_{\rm A}$, which greatly exceeds the disk scale height
$H$). We note in passing that a long effective
lever arm also characterizes angular momentum transport
by spiral waves in the disk.

The requirement that, in a steady state, the torque
exerted by a large-scale magnetic field at the surface of a thin and
nearly Keplerian accretion disk balances the inward 
angular momentum advection rate can be written as $\dot M_{\rm
acc}= 2 \, r^{5/2} \, |B_{\rm
z}\, B_{\rm \phi,s}|/(G\, M_*)^{1/2}$ (see eq. [17]).
Assuming a rough equality between the vertical and azimuthal
surface field components, this relation implies
that, at a distance of $1\ \au$ from a solar-mass protostar, a $1\ {\rm G}$
field (the value indicated by meteoritic data for the
protosolar nebula; Levy and Sonett 1978) could induce accretion
at a rate of $\sim 2 \times
10^{-6}\ \myr$, which is compatible with the mean values
inferred in embedded protostars (see Sec. II.B). For comparison,
the minimum-mass solar nebula model implies a thermal pressure at $1\
\au$ that corresponds to an equipartition magnetic field of $\sim 18\ {\rm
G}$, which implies that a $\sim 1\ {\rm G}$ field could be readily
anchored in the protosolar disk at that location.

\subsection{B.~~Wind-Driving Protostellar Disk Models}
The properties of protostellar accretion disks 
and of integrated \break
disk/wind systems have, so far, been
derived only under highly simplified assumptions. Among the
papers that can be consulted on this topic are K\"onigl (1989,
1997), Wardle and K\"onigl (1993), Ferreira and Pelletier
(1993a,b, 1995; also Ferreira 1997),
Lubow et al. (1994), Li (1995, 1996), and Reyes-Ruiz and Stepinski (1996).
A plausible origin for an open disk magnetic field
is the interstellar field that had
originally threaded the magnetically supported molecular cloud
and that was subsequently carried in by the collapsing core.
This picture is favored over disk-dynamo interpretations in view
of the fact that the latter typically produce closed (quadrupolar) field
configurations, although we note that scenarios for opening dynamo-generated
field lines have been considered in the literature (e.g., Tout and Pringle
1996; Curry, Pudritz, and Sutherland 1994). 

The theory and simulations of the previous sections assumed
ideal MHD.  While this approximation is adequate for the surface
layers of protostellar disks, the disk
interiors are typically weakly ionized, and their study
entails the application of multifluid MHD. For typical
parameters of disks around solar-mass YSOs, one can distinguish between the
low-density regime (on scales $r \simlt 100\ \au$), in which the
current in the disk is carried by metal ions and electrons (whose
densities are determined from the balance between ionizations by
cosmic rays and recombinations on grain surfaces), and the
high-density regime (on
scales $r \simlt 1-10$ AU), where the current is carried by small charged
grains or by ions and electrons that recombine without grains.
We restrict our attention to the low-density regime
and concentrate on the case where the
magnetic field ${\bf B}$ is
``frozen'' into the electrons and diffuses relative to the dominant
neutral component as a result of an ion--neutral drift (ambipolar diffusion).

To derive a self-consistent steady-state disk/wind configuration, one
combines the mass, momentum (radial, vertical, and angular), and energy
conservation relations, together with Maxwell's equations and
the generalized Ohm's law,
and imposes the requirements that the outflow pass through the
relevant critical points (see Sec. IV). So far only simple
prescriptions for the disk thermal structure (isothermal or
adiabatic) and conductivity (ambipolar diffusion in the density
regime where the ion density is constant or Ohmic diffusivity
parametrized using a
``turbulence'' prescription) have been considered. These are
probably adequate for obtaining the basic structure of the disk,
but more realistic calculations are needed to correctly model
the transition region between the disk and the wind. 

The vertical structure of a generic centrifugal wind-driving
disk (or ``active'' surface layer; see Sec. V.C) can be divided
into 3 distinct zones (see Fig. 3): a
quasi-hydrostatic region near the
midplane of the disk, where
the bulk of the matter is concentrated and most of the
field-line bending takes place, a transition zone where
the inflow gradually diminishes with height, and an outflow
region that corresponds to the base of the
wind. The first two regions are characterized by a radial inflow
and sub-Keplerian rotation, while the gas at the base of the
wind flows out with $V_{\phi} > V_{\rm K}$.
The boundary conditions for the stationary disk simulations 
discussed in Sec. IV.B arise from the physical
properties of this latter region.  

What determines the disk structure and how does the magnetic
field extract the angular momentum of the accreting gas?
The quasi-hydrostatic region is matter dominated, with the ionized plasma
and magnetic field being carried around by the neutral
material. The ions are braked by a magnetic torque, which is transmitted to the
neutral gas through the frictional (ambipolar-diffusion) drag; therefore
$V_{\rm i\phi}<V_\phi$ in this region (with the subscript
$i$ denoting ions). The neutrals thus lose angular momentum to the field,
and their back reaction leads to a buildup of the azimuthal
field component $|B_\phi|$ away from the midplane.
The loss of angular momentum enables the neutrals
to drift toward the center, and in doing so they exert a
radial drag on the field lines. This drag must be balanced by magnetic
tension, so the field lines bend away from the rotation axis.
This bending builds up the ratio $B_{\rm r}/B_{\rm z}$, which
needs to exceed $1/\sqrt{3}$ at the disk surface to launch a
centrifugally driven wind. The magnetic tension force, transmitted
through ion--neutral collisions, contributes to the radial support of the
neutral gas and causes it to rotate at sub-Keplerian speeds.

{\psfig{figure=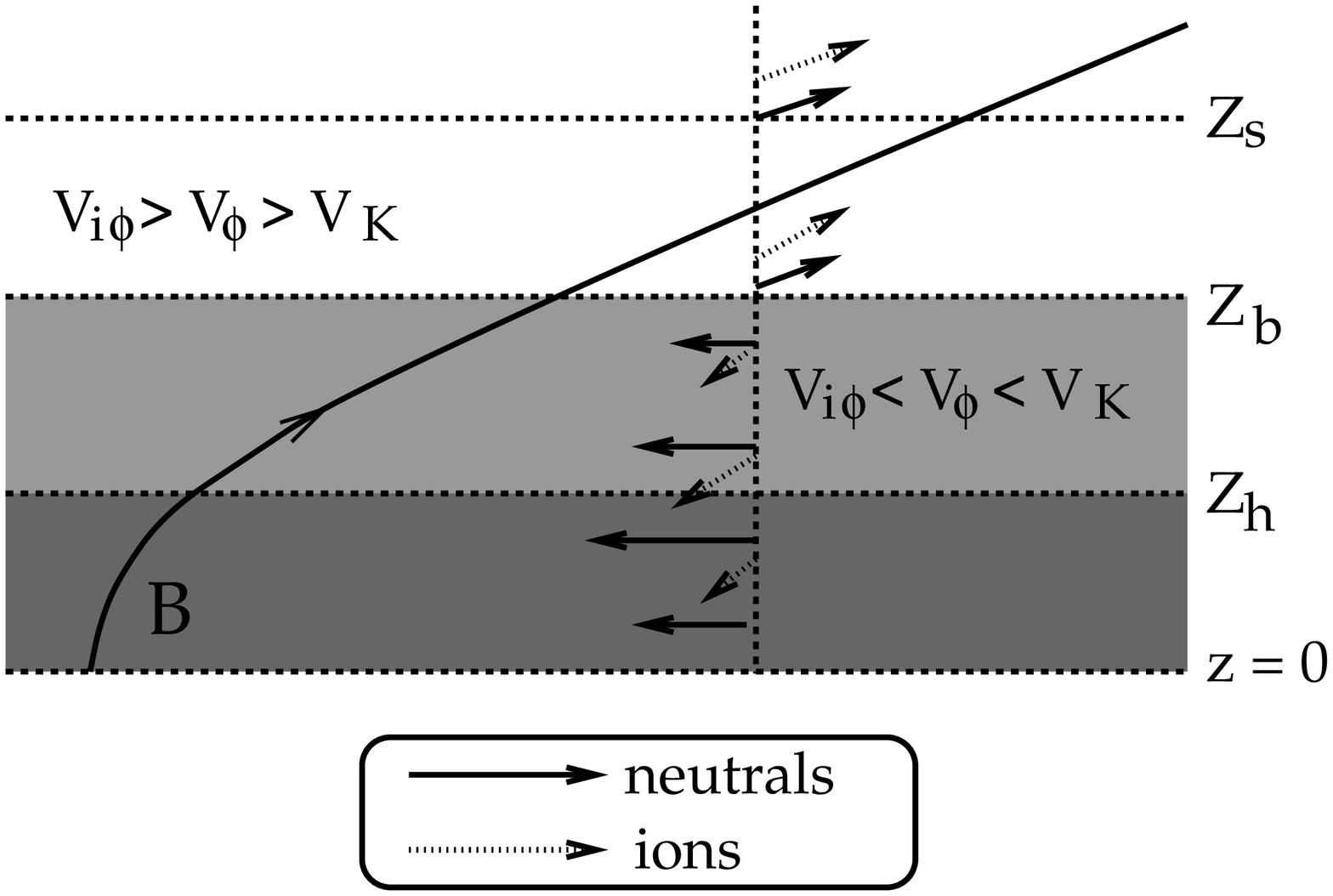,height=3.7in,width=4.2in} 
\caption{Figure 3.\capskip Schematic diagram of the vertical structure of an
ambipolar 
\break 
diffusion-dominated disk, showing a representative field line
and the
\break
poloidal velocities of the
neutral (solid arrowheads) and the ionized (open arrowheads)
fluid components.  Note that the poloidal velocity of
the ions vanishes at the midplane ($z=0$) and is small for both
fluids at the base of the wind ($z=Z_{\rm b}$). The relationship between the
azimuthal velocities is also indicated. 
}}

\bigskip
The growth of the radial and azimuthal field components on moving away
from the midplane results in a magnetic pressure gradient that tends to
compress the disk. The vertical compression by the combined magnetic and
tidal stresses is, in turn, balanced by the thermal pressure
gradient. The magnetic energy density becomes dominant
as the gas density decreases, marking the beginning of the transition 
zone (at $z=Z_{\rm h}$). The field above this point is nearly force free
($[{\bf \nabla \times B}] {\bf  \times B} \approx 0$), so the
field lines, which vary only on a length scale $\sim r$, are locally straight.
This is the basis for the adoption of force-free
initial field configurations in the   
simulations reviewed in Sec. IV.B.

The field angular velocity, given by $\omega =
(V_{\rm i \phi} \, - \, V_{\rm i z} B_{\phi} / B_{\rm z} )/r$,
is a field-line constant (corresponding to $\psi$ in the
ideal-MHD case; see Sec. III.A).
The ion angular velocity $V_{\rm i
\phi}/r$ differs somewhat from $\omega$ but still changes
only slightly along the field. Since the field lines bend away
from the symmetry axis, the cylindrical radius $r$, and
hence $V_{\rm i\phi}$, increase along any particular field line, whereas
$V_\phi$ decreases because of the near-Keplerian rotation law. Eventually a
point is reached where $(V_{\rm i \phi} - V_\phi)$ changes sign. At this point,
the magnetic stresses on the neutral gas are small and its angular velocity
is almost exactly Keplerian. Above this point, the field lines overtake the
neutrals and transfer angular momentum back to the matter, and
the ions start to push the neutrals out in both the radial
and the vertical directions.  It is thus natural to identify the
base $Z_{\rm b}$ of the wind with the location where the angular velocity
of the field lines becomes equal to the Keplerian angular
velocity. The mass outflow rate is fixed by the height
$Z_{\rm s}$ of the effective sonic point of the wind.

As was first recognized by Wardle and K\"onigl (1993), one can derive key
constraints on a viable solution by neglecting the vertical
component of the neutral velocity, which is generally a good
approximation throughout much of the disk column. This
simplifies the problem by transforming the radial and azimuthal
components of the neutral momentum equation into algebraic
relations. One important constraint that can be derived in this
way expresses the intuitively obvious condition that the
neutrals must be able to
couple to the magnetic field on
an orbital time scale if magnetic torques are to play a role in
the removal of angular momentum from the disk. In the
pure-ambipolar-diffusion regime, this condition is expressed by
the requirement that the neutral--ion coupling time
$1/\gamma \rho_{\rm i}$ (where $\gamma \approx
3.5 \times 10^{13}$ cm$^3$ g$^{-1}$ s$^{-1}$ is the collisional
coupling coefficient) be shorter than the dynamical
time. This is equivalent to requiring that the 
neutral--ion coupling parameter
$\eta \equiv \gamma \rho_{\rm i} r/ V_{\rm K}$ satisfy
$$
\eta > 1\, . \eqno(20)
$$
This condition is quite general and is relevant also to disk models
in which magnetic torques associated with small-scale magnetic
fields transfer angular momentum radially through the disk
(see chapter by Stone et al.; in the latter case eq. [20] identifies the
linearly unstable regime of the magnetorotational instability,
although evidently a much higher minimum value of $\eta$ is required for
significant nonlinear growth).

Additional parameter constraints that can be derived by using the
hydrostatic approximation for a wind-driving disk model in the
pure-ambipolar-diffusion regime are given by
$$
(2\eta)^{-1/2}\simlt a \simlt 2 \simlt  \epsilon \eta
\simlt V_{\rm K}/2 C_{\rm s}\, , \eqno(21)
$$
where $a \equiv B_0/(4 \pi \rho_0)^{1/2} C_{\rm s}$ is the ratio of the
midplane (subscript 0) Alfv\'en speed to the isothermal sound speed $C_{\rm
s}$ and $\epsilon \equiv -V_{r 0}/C_{\rm s}$ is the normalized
midplane inflow speed. The first inequality corresponds to the requirement
that the disk remain sub-Keplerian, the second
to the wind-launching condition at the disk surface (subscript
$b$) $B_{\rm r}(Z_{\rm b}) \simgt B_0/\sqrt{3}$, and the third to the
requirement that the base of the wind lie well above a density scale height
in the disk. The second and third inequalities together
imply that the vertical magnetic stress dominates the
gravitational tidal stress in confining the disk: this is a
generic property of this class of disk solutions that does not
depend on the nature of the magnetic diffusivity. The last
inequality expresses the requirement that the ambipolar diffusion heating rate
at the midplane not exceed the rate $\rho_0 |V_{r 0}| V_{\rm k}/2r$
of gravitational potential energy release.  In turn, $C_{\rm s}/V_{\rm
K}$ must be $\ll 1/(1+\epsilon )$ to guarantee that the
disk is in near-Keplerian motion and geometrically thin. One can
also place upper limits on the density at the sonic point to
ensure that the bulk of the disk material is hydrostatic and
that $\dot M_{\rm wind}$ does not exceed $\dot M_{\rm acc}$.
Similar constraints are derived by applying this analysis to
the other diffusivity regimes identified above. The solutions
that satisfy these constraints tend to have $\epsilon \simlt 1$
and $a \simgt 1$. Furthermore, the magnetic field in these
models, whose magnitude is essentially determined from the
condition that all the angular momentum liberated by the accreting matter is
transported by a centrifugally driven wind, automatically lies in a
``stability window'' where it is strong enough not to be affected
by the magnetorotational instability but not so strong as to be
subject to the radial interchange instability (K\"onigl and
Wardle 1996).

As we have noted, the steady-state disk models are useful for
guiding the choice of boundary conditions at the base of the outflow
in numerical simulations of winds from quasi-stationary disks.
In a complementary approach (which typically employs a highly
simplified treatment of the disk outflow and
therefore is not suitable for a detailed study of the disk wind),
one can generalize the steady-state models and investigate the
time evolution of wind-driving accretion disks (e.g. Lovelace et al. 1994;
K\"onigl 1997). The time-dependent
models  can be utilized to explore the feedback effect between
the magnetic flux distribution, which affects the angular
momentum transport in the disk, and the radial inflow induced by
the magnetic removal of angular momentum, which can modify the
flux distribution through field-line advection.

\subsection{C.~~Outflows from the Protostellar Vicinity}
As one moves to within a few AU from the protostar, the disk
hydrogen column density can become so
large ($\simgt 192\ {\rm g\ cm^{-2}}$) that cosmic rays are
excluded from the disk interior. (For reference, the column
density of the minimum-mass solar nebula model is $1.7 \times
10^3\ {\rm g\ cm^{-2}}$ at $1\ \au$). At
larger columns the ionization rate is dominated by the decay of radioactive
elements (notably $^{26}$Al if it is present and $^{40}$K;
Stepinski 1992), but the charge density generally becomes too
low for the magnetic field to be effectively coupled to the
matter. Consequently, the interior of
the disk becomes inert and magnetically mediated accretion can
only proceed through ``active'' surface layers that extend to
a depth of $\sim 96\ {\rm g\ cm^{-2}}$ on each side of the
disk. This scenario applies both to a wind-driving disk (Wardle
1997) and to the case where small-scale magnetic fields produce
an effective viscosity within the disk (Gammie 1996). Galactic
cosmic rays could reach the disk along the open magnetic field
lines that thread it, but a super-Alfv\'enic outflow along these
field lines would tend to exclude them. External ionization
could, however, also be effected by stellar X-rays (see chapter
by Glassgold et al.) as well as by fast particles accelerated
in stellar flares. In addition, heating by stellar irradiation
could contribute to the collisional ionization of the surface layers (e.g.,
D'Alessio et al. 1998).

The entire disk can recover an adequate coupling with the
magnetic field once the central temperature becomes large enough for
collisonal ionization to be effective. This first occurs when the
temperature increases above $\sim 10^3\, {\rm K}$ and potassium
is rapidly ionized (e.g., Umebayashi and Nakano 1981). For disks in
which magnetic fields dominate the angular momentum transport,
it may be possible that Joule dissipation could maintain the requisite
degree of ionization for efficient gas--field coupling
($\eta>1$; see eq. [20]).
Li (1996) first explored this possibility for the inner regions of
wind-driving protostellar disks and concluded that, for a disk that is in the
ambipolar-diffusion regime on scales $\simlt 1\ \au$, a
self-consistent model can only be constructed if the accretion
rate is very high ($\simgt 10^{-5}\ \myr$). 

A possible mechanism for recoupling the gas to the field in
the vicinity of the YSO is the thermal ionization
instability originally discussed in the context of dwarf novae
and more recently invoked as a possible explanation of FU Orionis
outbursts (e.g., Bell \& Lin 1994).  In this picture,
accretion in the innermost disk proceeds in a nonsteady fashion, with
a ``gate'' at $r \simlt 0.25$ AU opening every $\sim 10^3$ yr or so
after the accumulated
column density has become large enough to trigger the instability.  During the
``high'' phase of the instability (which lasts $\sim 10^2\ \yr$
and is identified with an outburst) the gas is hot ($T \simgt 10^4$ K) and
almost completely ionized, and mass rains in at a rate $\sim
(1-30) \times 10^{-5} \, M_{\odot} \, {\rm yr}^{-1}$, whereas
during the ``low''
phase the temperature and degree of ionization decline sharply, and
the accretion rate drops to $\sim (1-30) \times 10^{-8} \, M_{\odot} \,
{\rm yr}^{-1}$. In the context of the magnetized disk model,
one can attribute the increase in the
accretion rate as the gas becomes highly ionized to the
reestablishment of good coupling between the field and the matter,
which allows the field to extract the angular momentum of
the accreting gas. This could account both for the
marked increase in $\dot M_{\rm acc}$ and for the strong disk
outflow that accompanies it (see Sec. II.B). The magnetic
recoupling idea may be relevant to these outbursts even
if the wind is not the dominant angular momentum transport
mechanism in the disk provided that the viscosity has a magnetic
origin (see chapter by Stone et al.). We note, however, that
the relatively large ($\sim 0.1$) $\dot M_{\rm wind}/\dot M_{\rm
acc}$ ratio indicated in some of the outburst sources is consistent
with the bulk of the disk angular momentum being removed by the wind. 

At distances of a few stellar radii, the stellar magnetic field
could be strong enough to drive outflows from the disk. Various
scenarios have been considered in this connection, including
individual magnetic loops ejecting diamagnetic blobs through a
magnetic surface drag force (King and Regev 1994); a
steady-state configuration in which mass is transferred to the
star near the corotation radius and an outflow is driven along
adjacent (but disconnected from the star) field lines (Shu et
al. 1994), and time-dependent ejection associated with the
twisting, expansion, and subsequent reconnection of field lines
that connect the star with the disk (Hayashi et al. 1996;
Goodson et al. 1997). Numerical simulations are rapidly reaching
the stage where they could be utilized to identify the relevant
mechanisms and settle many of the outstanding questions. There
are already indications from some of the existing simulations that
magnetically channeled accretion from the disk to the star 
is most likely to occur when the disk carries a strong axial
magnetic field that reconnects
with the stellar field at an equatorial X-point; the resulting
field configuration appears to be quasi-steady and gives rise to sturdy
centrifugally driven outflows along open field lines (Hirose et
al. 1997; Miller and Stone 1997).

Stellar field-driven outflows are the subject of the
chapter by Shu et al. This class of models is based on the realization that
stellar magnetic field lines can be inflated and
opened up through an interaction with a surrounding disk, and that
a centrifugal wind can be driven out along the opened
field lines. In these scenarios, the outflow typically originates
near the inner disk radius, where the disk is truncated by the
stellar magnetic stresses. This contrasts with the scenarios
considered in this chapter, wherein a disk-driven wind
originates (and can contribute to the disk angular momentum
transport) over a significant range of radii. It is worth noting
in this connection that the massive inflows that
characterize FU Orionis outbursts are expected to crush the
respective stellar magnetospheres, so the strong outflows that are inferred to
originate from the circumstellar disks during these outbursts are unlikely to 
be driven along stellar field lines (Hartmann and Kenyon
1996). Coupled with the apparent inadequacy of thermal-pressure
and radiative driving, this argument provides strong support for
the relevance of disk-driven hydromagnetic winds that are not
associated with a stellar magnetic field to the observed
outflows from FU Orionis outburst sources. The ramifications of
this argument become even stronger if a significant fraction of the mass
accumulation in solar-type stars occurs through such outbursts
during the early phases of their evolution (Hartmann 1997).
To be sure, the ``distributed'' disk-wind models discussed in
this chapter are meant to apply to YSOs in general and not just
when they undergo an outburst, but only during an FU Orionis
outburst does the disk become sufficiently luminous that an
unambiguous observational signature of an extended disk wind
can be obtained.
  
\mainsection{{V}{I}.~~CONCLUSIONS}

Centrifugally driven winds from disks threaded by open magnetic
field lines provide the most efficient way of tapping the
gravitational potential energy liberated in the accretion
process to power an outflow. The fact that such winds
``automatically'' carry away angular momentum and thus
facilitate (and possibly even control) the accretion process
makes them an attractive explanation for the ubiquity of jets in
YSOs and in a variety of other accreting astronomical objects.
One of the key findings of recent numerical simulations of MHD
winds from disks is that such outflows are indeed easy to
produce and maintain under a variety of surface boundary
conditions. These simulations have also verified the ability of
such outflows to self-collimate and give rise to narrow jets, as
well as a variety of other characteristics that are consistent
with YSO observations. In the case of protostellar disks, the
presence of open field lines is a natural consequence of their formation
from the collapse of magnetically supported molecular cloud
cores. Although a stellar magnetic field that threads the disk
could in principle also play a similar role, this possibility is
unlikely to apply to the strong outflows associated with FU
Orionis outbursts. 
(More generally, it is worth noting that, in contrast to disk
field-driven outflows, a scenario that invokes a stellar
field may not represent a universal
mechanism since many cosmic jet sources are associated with a
black hole that does not provide an anchor for a central magnetic field.)
Much progress has also been achieved in constructing global MHD
disk/wind models, but the full elucidation of this picture and
its consequences for star formation remains a challenge for the future.
\bigskip
{\it Acknowledgments.}\/  We thank Vincent Mannings, Rachid Ouyed, and the
anonymous referee for helpful
comments on the manuscript. This work was supported in part by
NASA grant NAG 5-3687 (A.K.) and by an operating grant from the
Natural Science and Engineering Research Council of Canada
(R.P.).
\vfill\eject
\null

\vskip .5in
\centerline{\bf REFERENCES}
\vskip .25in

\ref{Appl, S., and Camenzind, M. 1992. The stability 
of current carrying jets.  
{\refit Astron.\ Astrophys. \/} 256:354--370.}

\ref{Appl, S., and Camenzind, M. 1993. The structure of MHD jets:
a solution to the non-linear Grad-Shafranov equation.  
{\refit Astron.\ Astrophys. \/} 274:699--706.}

\ref{Bacciotti, F., and Chiuderi, C. 1992. Axisymmetric
magnetohydrodynamic equations: exact solutions for stationary
incompressible flows. {\refit Phys.\ Fluids\ B\/} 4:35--43.}

\ref{Balbus, S. A., and Hawley, J. F. 1991. A powerful local
shear instability in weakly magnetized disks. I - Linear analysis. II - 
Nonlinear evolution. {\refit Astrophys.\ J.\/}  376, 214--233.}

\ref{Bell, K. R. 1994. Reconciling accretion scenarios with
inner holes: the thermal instabilioty and the 2$\mu$m gap.
In {\refit ASP Conf. Ser. 62, The Nature and
Evolution of Herbig Ae/Be Stars\/}, eds.\ P. S. Th\'e, M. R. P\'erez,
and E. P. J. van den Heuvel (San Francisco: ASP), pp.\ 215--218.}

\ref{Bell, K. R., and Lin, D. N. C. 1994. Using FU Orionis
outbursts to constrain self-regulated protostellar disk models.
{\refit Astrophys.\ J.\/} 427:987--1004.}

\ref{Bell, A. R., and Lucek, S. G. 1995.  
Magnetohydromagnetic jet formation.
{\refit Mon.\ Not.\ Roy.\ Astron.\ Soc.\/} 277:1327--1340.}

\ref{Blandford, R. D., and Payne, D. G. 1982.  Hydromagnetic
flows from accretion discs and the production of radio jets.
{\refit Mon.\ Not.\ Roy.\ Astron.\ Soc.\/} 199:883--903.}

\ref{Bontemps, S., Andr\'{e}, P., Terebey, S., and Cabrit, S. 1996.
Evolution of outflow activity around low-mass embedded young stellar objects.
{\refit Astron.\ Astrophys.\/} 311:858--872.}

\ref{Burrows, C. J., et al. 1996. Hubble Space Telescope
observations of the disk and jet of HH 30. {\refit Astrophys.\
J.\/} 473:437--451.}

\ref{Cabrit, S., and Andr\'e, P. 1991. An observational
connection between circumstellar disk mass and molecular outflows.
{\refit Astrophys.\ J.\/} 379:L25-L28.}

\ref{Cabrit, S., Edwards, S., Strom, S. E., and Strom, K. M.
1990. Forbidden line emission and infrared excesses in T Tauri
stars - Evidence for accretion-driven mass loss? {\refit
Astrophys.\ J.\/} 354:687--700.}

\ref{Cabrit, S., Raga, A. C., and Gueth, F. 1997. Models of
bipolar molecular outlows. In {\refit Harbig-Haro Flows and the
Birth of Low Mass Stars\/}, eds.\ B. Reipurth and C. Bertout
(Dordrecht: Kluwer), pp.\ 163--180.}

\ref{Calvet, N., Hartmann, L., and Kenyon, S. J. 1993. Mass loss
from pre--main-sequence accretion disks. I. The accelerating
wind of FU Orionis. {\refit Astrophys.\ J.\/} 402:623--634.}

\ref{Cao, X., and Spruit, H. C. 1994. Magnetically driven wind
from an accretion disk with low-inclination field lines.
{\refit Astron.\ Astrophys.\/} 287:80--86.}

\ref{Ciolek, G. E., and K\"onigl, A. 1998. Dynamical collapse of
nonrotating magnetic molecular cloud cores: evolution through
point-mass formation. {\refit Astrophys.\ J.\/} 504:257--.}

\ref{Clarke, D. A., Norman, M. L., and Burns, J. O. 1986.
Numerical simulations of a magnetically confined jet.
{\refit Astrophys.\ J.\/} 311:L63--L67.

\ref{Contopoulos, J. 1995. A simple type of magnetically driven jets: An 
astrophysical plasma gun. {\refit Astrophys.\ J.\/} 450:616--627.}

\ref{Contopoulos, J., and Lovelace, R. V. E. 1994. Magnetically
driven jets and winds: exact solutions. {\refit Astrophys.\ J.\/} 
429:139--152.}

\ref{Corcoran, M., and Ray, T. 1997. Forbidden emission lines in
Herbig Ae/Be stars. {\refit Astron.\ Astrophys.\/} 321:189-201.}

\ref{Corcoran, M., and Ray, T. 1998. Wind diagnostics and
correlations with the near-infrared excess in Herbig Ae/Be stars.
{\refit Astron.\ Astrophys.\/} 331:147-161.}

\ref{Crutcher, R.  M., Mouschovias, T. Ch., Troland, T. H., and
Ciolek, G. E. 1994. Structure and evolution of magnetically supported 
molecular clouds: evidence for ambipolar diffusion in the Barnard 1 cloud.
{\refit Astrophys.\ J.\/} 427:839--847.}

\ref{Crutcher, R. M., Roberts, D. A., Mehringer, D. M., and
Troland, T. H. 1996. H I Zeeman measurements of the magnetic field in 
Sagittarius B2. {\refit Astrophys.\ J.\/} 462:L79--82.}

\ref{Crutcher, R. M., Troland, T. H., Goodman, A. A., Heiles,
C., Kaz\'{e}s, I., and Myers, P. C. 1993. OH Zeeman observations
of dark clouds. {\refit Astrophys.\ J.\/} 407:175--184.}

\ref{Curry, C., Pudritz, R. E., and Sutherland, P. G. 1994.
On the global stability of magnetized accretion disks. I. 
Axisymmetric modes.
{\refit Astrophys.\ J.\/} 434:206--220.}

\ref{D'Alessio, P., Cant\'o, J., Calvet, N., and Lizano,
S. 1998. Accretion disks around young objects. I. The detailed
vertical structure. {\refit Astrophys.\ J.\/} 500:411--427.}

\ref{Eisl\"offel, J, and Mundt, R. 1997. Parsec-scale jets from
young stars. {\refit Astron.\ J.\/} 114:280--287.}
                
\ref{Ferreira, J. 1997. Magnetically-driven jets from Keplerian
accretion discs. {\refit Astron.\ Astrophys.\/} 319:340--359.}

\ref{Ferreira, J., and Pelletier, G. 1993a. Magnetized
accretion-ejection structures. I. General statements. {\refit
Astron.\ Astrophys.\/} 276:625--636.}

\ref{Ferreira, J., and Pelletier, G. 1993b. Magnetized
accretion-ejection structures. II. Magnetic channeling around compact objects.
{\refit Astron.\ Astrophys.\/} 276:637--647.}

\ref{Ferreira, J., and Pelletier, G. 1995. Magnetized
accretion-ejection structures. III. Stellar and extragalactic
jets as weakly dissipative disk outflows.
{\refit Astron.\ Astrophys.\/} 295:807--832.}

\ref{Gammie, C. F. 1996. Layered Accretion in T Tauri Disks.
{\refit Astrophys.\ J.\/} 457:355--362.}

\ref{Gomez, M., Whitney, B. A., and Kenyon, S. J. 1997. A survey
of optical and near-infrared jets in taurus embedded
sources. {\refit Astron.\ J.\/} 114:1138--1153.}

\ref{Goodson, A. P., Winglee, R. M., and B\"ohm,
K-.H. 1997.\break Time-dependent accretion by magnetic young stellar
objects as a launching mechanism for stellar jets. {\refit
Astrophys.\ J.\/} 489:199--209.}

\ref{Greaves, S., and Holland, W. S. 1998. Twisted magnetic
field lines around protostars. {\refit Astron.\ Astrophys.\/} 333:L23--L26.}

\ref{Greaves, J. S., Holland, W. S., and Murray,
A. G. 1995. Magnetic field compression in the Mon R2 cloud core.
{\refit Mon.\ Not.\ Roy.\ Astron.\ Soc.\/} 297:L49--L52.}

\ref{Greaves, J. S., Murray, A. G., and Holland,
W. S. 1994. Investigating the magnetic field structure around
star formation cores. {\refit Mon.\ Not.\ Roy.\ Astron.\ Soc.\/}
284:L19--L22.} 

\ref{Hartigan, P., Edwards, S., and Ghandour, L. 1995. Disk
accretion and mass loss from young stars. {\refit Astrophys.\
J.\/} 452:736--768.}

\ref{Hartigan, P., Morse, J., and Raymond, J. 1994. Mass-loss
rates, ionization fractions, shock velocities, and magnetic
fields of stellar jets. {\refit Astrophys.\ J.\/} 136:124--143.}

\ref{Hartmann, L. 1997. The observational evidence for
accretion. In {\refit Harbig-Haro Flows and the
Birth of Low Mass Stars\/}, eds.\ B. \break
Reipurth and C. Bertout (Dordrecht: Kluwer), pp.\ 391--405.}

\ref{Hartmann, L., and Calvet, N. 1995. Observational
constraints on Fu Ori winds. {\refit Astron.\ J.\/} 109:1846--1855.}

\ref{Hartmann, L., and Kenyon, S. J. 1996. The FU Orionis Phenomenon.
{\refit Ann.\ Rev.\ Astron.\ Astrophys.\/} 34:207--240.}

\ref{Hartmann, L., Kenyon, S. J. and Calvet, N. 1993. The excess
infrared emission of Herbig Ae/Be stars: disks or envelopes?
{\refit Astrophys.\ J.\/} 407:219--231.}

\ref{Hayashi, M. R., Shibata, K., and Matsumoto, R. 1996. X-ray
flares and mass outflows driven by magnetic interaction between
a protostar and its surrounding disk. {\refit Astrophys.\ J.\/} 468:L37--L40.}

\ref{Heinemann, M., and Olbert, S. 1978.
Axisymmetric ideal MHD stellar wind flow.
{\refit J.\ Geophys.\ Res.\/} 83:2457--2460.}

\ref{Heyvaerts, J., and Norman, C. A. 1989.       
The collimation of magnetized winds.
{\refit  Astrophys.\ J.\/} 347:1055--1081.}

\ref{Heyvaerts, J., and Norman, C. A. 1997. In 
{\refit IAU Symposium 182, Herbig-Haro Flows and the Birth
of Low Mass Stars \/}, eds.\ B. Reipurth, and C. Bertout 
(Dordrecht: Kluwer), pp.\ 275--290.} 

\ref{Hildebrand, R. H., Dotson, J. L., Dowell,
C. D., Platt, S. R., Schleuning, D., Davidson, J. A., and Novak,
G. 1995. Far-infrared polarimetry.
In {\refit ASP Conf. Ser. 73, Airborne Astronomy Symposium on the Galactic
Ecosystem\/}, eds.\ M. R. Haas, J. A. Davidson, and E. F. Erickson
(San Francisco: ASP), pp.\  97--104.}

\ref{Hillenbrand, L. A., Strom, S. E., Vrba, F. J., and Keene,
J. 1992. Herbig Ae/Be stars - Intermediate-mass stars surrounded
by massive circumstellar accretion disks. {\refit Astrophys.\
J.\/} 397:613-643.}

\ref{Hirose, S., Uchida, Y., Shibata, K., and Matsumoto,
R. 1997. {\refit Publ.\ Astron.\ Soc.\ Japan\/} 49:193--205.}

\ref{King, A. R., and Regev, O. 1994. Spin rates and mass loss in
accreting T Tauri stars. {\refit Mon.\ Not.\ Roy.\ Astron.\
Soc.\/} 268:L69--L73.}

\ref{K\"onigl, A. 1989. Self-similar models of magnetized accretion disks.
{\refit Astrophys.\ J.\/} 342:208--223.}

\ref{K\"onigl, A. 1997. Magnetized accretion disks and the
origin of bipolar outflows. In {\refit ASP Conf. Ser. 121,
Accretion phenomena and
related outflows\/}, eds.\ D. T. Wickramasinghe, G. V. Bicknell,
and L. Ferrario (San Francisco: ASP), pp.\ 551--560.}

\ref{K\"onigl, A. 1999. Theory of bipolar outflows from
high-mass young stellar objects. {\refit New\ Astron.\ Rev.\/},
in press.}

\ref{K\"onigl, A., and Ruden, S. P. 1993. Origin of outflows and
winds. In {\refit Protostars \& Planets {I}{I}{I}\/}, eds.\ E. H. Levy and
J. I. Lunine (Tucson: Univ.\ of Arizona Press), pp.\ 641--688.}

\ref{K\"onigl, A., and Wardle, M. 1996. A comment on the stability of magnetic
wind-driving accretion discs. {\refit Mon.\ Not.\ Roy.\ Astron.\ Soc.\/} 
279:L61--L64.}

\ref{K\"ossel, D., M\"uller, E., and Hillebrandt, W. 1990.
Numerical simulations of axially symmetric magnetized jets.
{\refit Astron.\ Astrophys.\/} 229:401--415.}

\ref{Lada, C. J. 1985. Cold outflows, energetic winds, and enigmatic jets 
around young stellar objects. {\refit Ann.\ Rev.\ Astron.\
Astrophys.\/}\break
23:267--317.}

\ref{Lery, T., Heyvaerts, J., Appl, S., and Norman,
C. A. 1998. Outflows from magnetic rotators. I. Inner structure.
{\refit Astron.\ Astrophys.\/} 337:603--624.}

\ref{Levreault, R. M. 1988. Molecular outflows and mass loss in the 
pre-main-sequence stars. {\refit Astrophys.\ J.\/} 330:897--910.}

\ref{Levy, E. H., and Sonnett, C. P. 1978. Meteorite magnetism
and early solar system magnetic fields. In {\refit Protostars \&
Planets\/}, ed.\ T. Gehrels (Tucson: Univ.\ of Arizona Press), pp.\ 516--532.}

\ref{Li, Z.-Y. 1995. Magnetohydrodynamic disk-wind connection:
self-similar solutions. {\refit Astrophys.\ J.\/} 444:848--860.}

\ref{Li, Z.-Y. 1996. Magnetohydrodynamic disk-wind connection:
magnetocentrifugal winds from ambipolar diffusion-dominated accretion disks.
{\refit Astrophys.\ J.\/} 465:855--868.}

\ref{Lind, K. R., Payne, D. G., Meier, D. L., and Blandford, R. D. 1989.
Numerical simulations of magnetized jets.
{\refit Astrophys.\ J.\/} 344:89--103.}

\ref{Lovelace, R. V. E., Romanova, M. M., and Newman,
W. I. 1994. Implosive accretion and outbursts of active galactic
nuclei. {\refit Astrophys.\ J.\/} 437:136--143.}

\ref{Lubow, S. H., Papaloizou, J. C. B., and Pringle, J. 1994.
Magnetic field dragging in accretion discs. {\refit Mon.\ Not.\
Roy.\ Astron.\ Soc.\/} 267:235-240.}

\ref{Lynden-Bell, D., and Boily, C. 1994.
Self-similar solutions up to flashpoint in highly wound magnetostatics.
{\refit Mon.\ Not.\ Roy.\ Astron.\ Soc.\/} 267:146--152.}

\ref{Masson, C. R., and Chernin, L. M. 1993.
Properties of jet-driven molecular outflows.
{\refit Astrophys.\ J.\/} 414:230--241.}

\ref{Matsumoto, R., Uchida, Y., Hirose, S., Shibata, K., Hayashi, M. R.,
Ferrari, A., Bodo, G., and Norman, C. 1996.
Radio jets and the formation of active galaxies: accretion
avalanches on the torus by the effect of a large-scale magnetic field.
{\refit Astrophys.\ J.\/} 461:115--126.}

\ref{McKee, C. F., Zweibel, E. G., Goodman, A. A., and Heiles,
C. 1993. Magnetic fields in star-forming regions: theory.
In {\refit Protostars \& Planets {I}{I}{I}\/}, eds.\ E. H. Levy and
J. I. Lunine (Tucson: Univ.\ of Arizona Press), pp.\ 327--366.}

\ref{Meier, D., Edgington, S., Godon, P., Payne, D., and Lind, K. 1997.
A magnetic switch that determines the speed of astrophysical jets.
{\refit Nature\/} 388:350--352.}

\ref{Mestel, L. 1968. Magnetic braking by a stellar wind.
{\refit Mon.\ Not.\ Roy.\ Astron.\ Soc.\/} 138:359--391.}

\ref{Michel, F. C. 1969. Relativistic stellar-wind torques.
{\refit Astrophys.\ J.\/} 158:727--738.}

\ref{Miller, K. A., and Stone, J. M. 1997. Magnetohydrodynamic
simulations of stellar magnetosphere--accretion disk
interaction. {\refit Astrophys.\ J.\/} 489:890--902.}

\ref{Miroshnichenko, A., Ivezi\'c, {\v Z}., and Elitzur,
M. 1997. On protostellar disks in Herbig Ae/Be stars. {\refit
Astrophys.\ J.\/} 475:L41--L44.}

\ref{Mouschovias, T. Ch. 1991. Cosmic magnetism and the basic
physics of the early stages of star formation. In {\refit The
Physics of Star Formation and Early Stellar Evolution\/}, eds.\
C. J. Lada and N. D. Kylafis (Dordrecht: Kluwer), pp.\ 61--122.}

\ref{Mundt, R., and Ray, T. M. 1994. Optical outflows from
Haerbig Ae/Be stars and other high luminosity young stellar
objects. In {\refit ASP Conf. Ser. 62, The Nature and
Evolution of Herbig Ae/Be Stars\/}, eds.\ P. S. Th\'e, M. R. P\'erez,
and E. P. J. van den Heuvel (San Francisco: ASP), pp.\ 237--252.}

\ref{Ogilvie, G. I., and Livio, M. 1998.
On the difficulty of launching an outflow from an accretion disk.
{\refit Astrophys.\ J.\/} 499:329--339.}

\ref{Ostriker, E. 1998.
Self-similar magnetocentrifugal disk winds with cylindrical asymptotics.
{\refit Astrophys.\ J.\/} 486:291--306.}

\ref{Ouyed, R., and Pudritz, R. E. 1997.
Numerical simulations of astrophysical jets from Keplerian
accretion disks. I. Stationary models.
{\refit Astrophy.\ J.\/} 482:712--732.}

\ref{Ouyed, R., and Pudritz, R. E. 1997.
Numerical simulations of astrophysical jets from Keplerian
accretion disks. II. Episodic outflows.
{\refit Astrophy.\ J.\/} 484:794--809.}

\ref{Ouyed, R., and Pudritz, R. E. 1998.
Numerical simulations of astrophysical jets from Keplerian
accretion disks. III The effects of mass loading.
{\refit Mon.\ Not.\ Roy.\ Astron.\ Soc.\/} in press.}

\ref{Ouyed, R., Pudritz, R. E., and Stone, J. M. 1997.
Episodic jets from black holes and protostars.
{\refit Nature\/} 385:409--414.}

\ref{Pelletier, G., and Pudritz, R. E. 1992. Hydromagnetic disk
winds in young stellar objects and active galactic
nuclei. {\refit Astrophys.\ J.\/} 394:117--138.}

\ref{Pezzuto, S., Strafella, F., and Lorenzetti,
D. 1997. On the circumstellar matter distribution around Herbig
Ae/Be stars. {\refit Astrophy.\ J.\/} 485:290--307.}

\ref{Pudritz, R. E., and Norman, C. A. 1983.
Centrifugally driven winds from contracting molecular disks.
{\refit Astrophys.\ J.\/} 274:677--697.}

\ref{Pudritz, R. E., and Ouyed, R. 1997. 
Numerical simulations of jets from accretion disks.
In {\refit IAU Symposium 182: Herbig-Haro Flows and the Birth
of Low Mass Stars \/}, eds.\ B. Reipurth, and C. Bertout
(Dordrecht: Kluwer), pp.\ 259--274.}

\ref{Pudritz, R. E., Pelletier, G., and Gomez de Castro, A.
I. 1991. The physics of disk winds. In {\refit The Physics of
Star Formation and Early Stellar Evolution\/}, eds.\ C. J. Lada
and N. D. Kylafis (Dordrecht: Kluwer), pp.\ 539--564.}

\ref{Pudritz, R. E., Wilson, C. D., Carlstrom, J. E., Lay,
O. P., Hills, R. E., and Ward-Thompson, D. 1996. Accretions
disks around Class 0 protostars: the case of VLA 1623. {\refit Astrophys.\
J.\/} 470:L123--L126.}

\ref{Ray, T., Muxlow, T. W. B., Axon, D. J., Brown, A.,
Corcoran, D., Dyson, J., and Mundt, R. 1997. Large-scale
magnetic fields in the outflow from the young stellar object T Tauri S.
{\refit Nature\/} 385:415--417.}

\ref{Reipurth, B. 1991. Observations of Herbig-Haro objects. In
{\refit Low Mass Star Formation and Pre-Main Sequence
Objects\/}, ed. B. Reipurth (Munich: ESO), pp.\ 247--279.}

\ref{Reyes-Ruiz, M., and Stepinski, T. F. 1996. Axisymmetric
two-\break
dimensional computation of magnetic field dragging in accretion disks.
{\refit Astrophys.\ J.\/} 459:653--665.}

\ref{Romanova, M. M., Ustyugova, G. V., Koldoba, A. V., Chechetkin, V. M.,
and Lovelace, R. V. E. 1997.
Formation of stationary magnetohydrodynamic outflows from a disk
by time-dependent simulations.
{\refit Astrophys.\ J.\/} 482:708--711.}

\ref{Sauty, C., and Tsinganos, K. 1994. Nonradial and
nonpolytropic astrophysical outflows III. A criterion for the
transition from jets to winds. {\refit Astron.\ Astrophys.\/} 287:893--926.}

\ref{Schatzman, E. 1962. A theory of the role of magnetic
activity during star formation. {\refit Ann.\ Astrophys.\/} 25:18--29.}

\ref{Schleuning, D. A. 1998. 
Far-infrared and submillimeter polarization of 
OMC-1: evidence for magnetically regulated star
formation. {\refit Astrophys.\ J.\/} 493:811--825.}

\ref{Shibata, K., and Uchida, Y. 1986.
A magnetohydrodynamical mechanism for the formation of astrophysical
jets.II - Dynamical processes in the accretion of magnetized mass
in rotation.
{\refit Publ.\ Astron.\ Soc.\ Japan\/} 38:631--660.} 

\ref{Shu, F. H., Najita, J., Ostriker, E., Wilkin, F.,
Ruden, S., and Lizano, S. 1994. Magnetocentrifugally driven
flows from young stars and disks. I. A generalized
model. {\refit Astrophys.\ J.\/} 429:781--796.}

\ref{Skinner, S. L., Brown, A., and Stewart,
R. T. 1993. A high-sensitivity survey of radio continuum
emission from Herbig Ae/Be stars. {\refit Astrophys.\ J.\
Suppl.\/} 87:217--265.}

\ref{Spruit, H. C. 1996.
Magnetohydrodynamic jets and winds from accretion disks.
In {\refit NATO ASI Ser. C. 477, Evolutionary processes in 
binary stars. \/} eds. R. A. M. J. Wijers, M. B. Davies,
and C. A. Tout (Dordrecht: Kluwer), pp.\ 249--286.}   

\ref{Stepinski, T. F. 1992. Generation of dynamo magnetic fields in the 
primordial solar nebula. {\refit Icarus\/} 97:130--141.}

\ref{Stone, J. M., and Norman, M. L. 1992.
ZEUS-2D: a radiation magnetohydrodynamics code for astrophysical flows
in two space dimensions. II. The magnetohydrodynamic algorithms and tests.
{\refit Astrophys.\ J.\ Suppl.\/} 80:791--818.}

\ref{Stone, J. M., and Norman, M. L. 1994.
Numerical simulations of magnetic accretion disks.
{\refit Astrophys.\ J.\/} 433:746--756.}

\ref{Tout, C. A., and Pringle, J. E. 1996. Can a disc dynamo
generate large-scale magnetic fields? {\refit Mon.\ Not.\ Roy.\
Astron.\ Soc.\/} 281:219--225.}

\ref{Tsinganos, K., and Trussoni, E. 1990.
Analytic studies of collimated winds I. Topologies of 2-D 
helicoidal hydrodynamic solutions.
{\refit Astron.\ Astrophys.\/} 231:270--276.}

\ref{Uchida, Y., and Shibata, K. 1985.
A magnetohydrodynamic mechanism for the formation of astrophysical
jets. I - Dynamical effects of the relaxation of nonlinear
magnetic twists.
{\refit Publ.\ Astron.\ Soc.\ Japan\/} 37:31--46.}

\ref{Umebayashi, T., and Nakano, T. 1981.
Fluxes of energetic particles and the ionization rate in 
very dense interstellar clouds.
{\refit Publ.\ Astron.\ Soc.\ Japan\/} 33:617--635.}

\ref{Ustyugova, G. V., Koldoba, A. V., Romanova, M. M., Chechetkin, V. M.,
and Lovelace, R. V. E. 1995.
Magnetohydrodynamic simulations of outflows from accretion disks.
{\refit Astrophys.\ J.\/} 439:L39--L42.}

\ref{Wardle, M. 1997. Magnetically driven winds from
protostellar disks. In {\refit ASP Conf. Ser. 121, Accretion
Phenomena and Related Outflows\/}, eds. D. T. Wickramasinghe,
G. V. Bicknell, and L. Ferrario (San Francisco: ASP), pp.\
561--565.}

\ref{Wardle, M., and K\"onigl, A. 1993. The structure of
protostellar accretion disks and the origin of bipolar flows.
{\refit Astrophys.\ J.\/} 410:218--238.}

\ref{Weber, E. J., and Davis, L. 1967. The angular momentum of the
solar wind. {\refit Astrophys.\ J.\/} 148:217--227.}

\bye